\documentclass[%
reprint,
twocolumn,
amsmath,
amssymb,
aps,
prb,
nofootinbib,
superscriptaddress, 
floatfix,
longbibliography]{revtex4-2}

\usepackage{newtxtext, newtxmath}

\usepackage[pdftex]{graphicx}
\usepackage{hyperref}

\usepackage{bm}
\usepackage{braket}
\usepackage{amsfonts,mathrsfs}
\usepackage{mathtools}
\usepackage{xspace}
\usepackage{array}
\usepackage{multirow}
\usepackage{comment}

\hypersetup{colorlinks = true, urlcolor = blue, linkcolor = blue, citecolor = blue}

\makeatletter

\let\MYcaption\@makecaption
\makeatother

\usepackage{subcaption}
\makeatletter
\let\@makecaption\MYcaption
\makeatother

\usepackage{color}
\usepackage[normalem]{ulem} 

\newcommand{\bk}{\bm{k}}

\begin{document}

\title{Quantum geometric magnetic monopole and two-phase superconductivity in CeRh$_2$As$_2$}

\author{Kosuke Nogaki}
\email[]{kosuke.nogaki.scphys@niigata-u.ac.jp}
\affiliation{%
  Department of Physics, Kyoto University, Kyoto 606-8502, Japan
}%
\affiliation{%
RIKEN Center for Emergent Matter Science (CEMS), Wako 351-0198, Japan
}%
\affiliation{%
Department of Physics, Niigata University, Niigata 951-2181, Japan 
}
\affiliation{%
Institute for Research Administration, Niigata University, Niigata 951-2181, Japan
}

\author{Youichi Yanase}
\affiliation{%
  Department of Physics, Kyoto University, Kyoto 606-8502, Japan
}%

\date{\today}

\begin{abstract}
Recent angle-resolved photoemission spectroscopy (ARPES) and density functional theory plus Hubbard $U$ (DFT+$U$) studies revealed that a heavy-fermion superconductor CeRh$_2$As$_2$ exhibits van Hove singularities and the Dirac point near the Fermi level $E_{\mathrm F}$, which are key signatures of strong-correlation effects and quantum geometry.  
We have constructed a two-dimensional 12-orbital \textit{Dirac-Anderson} model as an effective model for CeRh$_2$As$_2$.  
The band structure and Fermi-surface topology of the Dirac-Anderson model agree well with the ARPES data and the DFT+$U$ calculations.  
We show that the quantum geometry strongly favors magnetic-monopole fluctuations because of the Dirac point at the $M$ point.  
By solving the linearized \'{E}liashberg equation, we demonstrate that the $B_{1u}$ and $B_{2g}$ representations, spin-triplet states originating from the Dirac point, exhibit the leading superconducting instabilities.  
By comparing the random-phase approximation and the fluctuation-exchange approximation, we further demonstrate that strong-correlation effects mitigate the influence of quantum geometry.
The phase diagram of CeRh$_2$As$_2$ under pressure is discussed in connection with the theoretical results.
\end{abstract}

\maketitle

\section{Introduction}
Topology, as well as symmetry breaking, has become a central paradigm in condensed matter physics for the classification of phases of matter~\cite{Wen2007textbook,Fradkin2013textbook}. 
Topological phases manifest in a wide range of systems, including integer/fractional quantum Hall states~\cite{Laughlin1981prb,Thouless1982prl}, quantum spin liquids~\cite{Savary2017rpp,Zhou2017rmp,Broholm2020science}, and topological insulators/superconductors~\cite{Qi2011rmp,Tanaka2012jpsj,Sato2016jpsj,Sato2017rpp}. 
A hallmark consequence of topological order is the bulk--edge correspondence. 
Notably, Majorana fermions, potential building blocks for fault-tolerant quantum computation, appear at the edges and defects of topological superconductors~\cite{Kitaev2001pu,Nayak2008rmp}.

Strongly correlated superconductors offer a natural platform for realizing topological superconductivity, since strong spin and multipole fluctuations, induced by electron correlations, often mediate unconventional pairing states~\cite{Moriya2000ap,Yanase2003pr}. 
Among these pairing states, spin-triplet odd-parity superconductors constitute particularly promising candidates for topological superconductivity~\cite{Aoki2022jpcm}. 
Nevertheless, conclusive examples of spin-triplet superconductors remain scarce, motivating the search for new materials and mechanisms.

To address the aforementioned issue, spin-singlet odd-parity superconductivity that exploits sublattice degrees of freedom has been proposed~\cite{Fischer2011prb,Maruyama2012jpsj,Yoshida2012prb,Fischer2023arcmp}. 
The locally noncentrosymmetric structure activates the sublattice degrees of freedom and staggered antisymmetric spin-orbit coupling.
Combining spin-orbit coupling with an external magnetic field stabilizes a sublattice-staggered odd-parity state through the spin-singlet pairing channel.
However, this phase remains fragile with respect to the orbital depairing effect~\cite{Mockli2018prb,Mockli2021prb}.
Therefore, large spin-orbit coupling and a sizable Maki parameter~\cite{Maki1964ppf} are essential, and $f$-electron heavy-fermion systems constitute suitable platforms.

CeRh$_2$As$_2$, a recently discovered locally noncentrosymmetric heavy-fermion superconductor, exhibits a two-phase $H$-$T$ superconducting phase diagram that has attracted considerable interest~\cite{Khim2021science,Kimura2021prb,Onisi2022fem,Kibune2022prl,Hafner2022prx,Kitagawa2022jpsj,Landaeta2022prx,Mishra2022prb,Ogata2023prl,Semeniuk2023prb,Siddiquee2023prb,Ogata2023npsm,Christovam2024prl,Chajewski2024prl,Chen2024prx,Chen2024prb,Wu2024cpl,Pfeiffer2024prl,Semeniuk2024prb,Ogata2024prb,Chen2024prl,Khim2021science}. 
The compound crystallizes in the space group $P4/nmm$ (No.\,129), which is nonsymmorphic and tetragonal and contains two inequivalent Ce sites per unit cell [Fig.~\ref{fig:band}(a)].
With a magnetic field applied along the $c$ axis, the system undergoes a transition from a low-field superconducting state to a high-field state.
The sublattice degrees of freedom encoded in the crystal structure, together with $f$-electron-driven superconductivity, strongly suggest that the high-field phase realizes an odd-parity superconducting state, as proposed in Ref.~\cite{Yoshida2012prb}.
Therefore, topological superconductivity is expected in CeRh$_2$As$_2$.
Indeed, our previous first-principles band-structure calculations showed that the high-field phase hosts topological crystalline superconductivity~\cite{Nogaki2021prr,Ishizuka2024prb}.

In addition, the Kondo coherence temperature of $T_{\mathrm{coh}}\!\sim\!30\,\mathrm{K}$, the large electronic specific heat coefficient of $\gamma\!\sim\!1000\,\mathrm{mJ\,mol^{-1}\,K^{-2}}$, and pronounced non-Fermi-liquid behaviors all point to strong-correlation effects that should be considered when discussing superconductivity and related phenomena in CeRh$_2$As$_2$~\cite{Khim2021science,Hafner2022prx,Mishra2022prb,Pfeiffer2024prl}.  
Such correlation effects have been examined within simplified tight-binding models~\cite{Nogaki2022prb,Lee2025prl}; however, because magnetic fluctuations and unconventional pairing are highly sensitive to the detailed electronic structure, a study based on a realistic and material-specific Hamiltonian remains indispensable.

Recent angle-resolved photoemission spectroscopy (ARPES) measurements have partially clarified the electronic structure of CeRh$_2$As$_2$, revealing a quasi-two-dimensional band structure, pronounced Fermi surface nesting, and a van Hove singularity~\cite{Chen2024prx,Chen2024prb,Wu2024cpl}.  
The quasi-two-dimensional character has also been suggested by nuclear magnetic resonance (NMR) and nuclear quadrupole resonance (NQR) studies~\cite{Kibune2022prl,Kitagawa2022jpsj,Ogata2023prl,Ogata2023npsm,Ogata2024prb} and neutron scattering experiments~\cite{Chen2024prl}.  
Both Fermi surface nesting and van Hove singularities are widely recognized as key features of strongly correlated electron systems~\cite{Yanase2003pr}.  
The density functional theory plus Hubbard $U$ (DFT+$U$) calculations show all the observed characteristics of the electronic band structure and further predict a $f$-electron Dirac point~\cite{Chen2024prx,Ishizuka2024prb}.  
Because Dirac points strongly enhance the quantum metric, a geometric quantity that quantifies the distance between adjacent quantum states in the Hilbert space of Bloch wave functions~\cite{Resta2011}, CeRh$_2$As$_2$ provides a fertile platform for investigating the interplay between quantum geometry and strong-correlation effects.

The quantum geometric tensor characterizes the geometry of the Hilbert space~\cite{Torma2022nrp,Liu2024nsr}. 
Its imaginary part is the Berry curvature, whereas the real part is known as the quantum metric.  
The former underlies a wide variety of topological phases.  
The quantum metric has recently become the focus of intense interest because of its potential applications in optical and transport responses~\cite{Gao2014,Sodemann2015prl,Du2018prl,Du2019nc,Ma2019nature,Wang2023,Gao2023,Han2024,annualrevOrenstein,Morimoto_JPSJreview,Nagaosa-Yanase}, flat-band superconductivity~\cite{Peotta2015nc,Julku2016prl,Liang2017prb, Torma2022nrp,Tian2023,Tanaka2025-sy,Banerjee2025-ew}, and fractional Chern insulators~\cite{Tang2011prl,Kai2011prl,Neupert2011prl}.  
Beyond these contexts, connection between quantum geometry 
and unusual electronic order has recently emerged as a research frontier~\cite{Kitamura2024prl,Alicea2025,jahin2025enhancedkohnluttingertopologicalsuperconductivity,heinsdorf2025altermagnetic,Kudo2025arxiv}.

Multipole physics is a fundamental framework for classifying atomic degrees of freedom~\cite{Sugano1970textbook}.
Extending atomic multipole to unit-cell degrees of freedom, the so-called \emph{augmented-multipole} theory, constitutes a rapidly developing research area~\cite{Watanabe2018prb,Hayami2018prb,Yatsushiro2021prb} that can provide a framework for describing a wide range of electronic order. 
Furthermore, formulating multipole moments in crystalline systems has uncovered an intimate connection with quantum geometry~\cite{Gao2018spin,Shitade2018prb,Gao2018orbital,Shitade2019prb,Daido2020prb,Kitamura2021prb}.  
In CeRh$_2$As$_2$, magnetic order develops inside the superconducting phase; NQR measurements reveal an unusual cancellation of internal fields at selected atomic sites, pointing to an odd-parity augmented multipole order~\cite{Kibune2022prl,Ogata2023prl,Ogata2024prb}.
Hence, CeRh$_2$As$_2$ is positioned at the intersection of research areas of superconductivity, strong-correlation effects, quantum geometry, and multipole physics.

The rest of this paper is organized as follows. 
In Sec.~\ref{sec:model}, we describe the 12-orbital Dirac-Anderson model of CeRh$_2$As$_2$.
We then discuss multipole fluctuations in Sec.~\ref{sec:suscep}.
The ferroic magnetic monopole fluctuation induced by quantum geometry is presented.
In Sec.~\ref{sec:sc}, superconducting properties are analyzed. 
The spin-triplet superconductivity concentrated on the Dirac band is revealed.
In Sec.~\ref{sec:flex}, the effect of self-energy correction on quantum geometry and magnetic anisotropy is explored. An implication for the superconducting phase diagram of CeRh$_2$As$_2$ is discussed.
Finally, Sec.~\ref{sec:summary} summarizes this paper.

\section{Model}
\label{sec:model}
The DFT+$U$ calculations have shown that a finite Coulomb interaction~$U$ transforms the three-dimensional band structure obtained at~$U=0$~\cite{Nogaki2021prr} into a quasi-two-dimensional one, in good agreement with the ARPES data~\cite{Chen2024prx,Ishizuka2024prb}.  
By means of finite Coulomb interaction~$U$, the unoccupied Ce-4$f$ bands shift upward, while the occupied band shifts downward and intersects $E_{\mathrm F}$~\cite{Ishizuka2024prb}.
This modification supports the heavy-fermion behavior of CeRh$_2$As$_2$~\cite{Khim2021science}.
The electronic band structure around the Fermi level~$E_{\rm F}$ mainly consists of Ce-4$f$, Rh1-5$d$, and Rh2-5$d$ electrons.
The ARPES measurement~\cite{Chen2024prx} and the DFT+$U$~\cite{Ishizuka2024prb} calculation have revealed the van Hove singularities at the $X$ point whose energy levels are $-75\pm60\,\mathrm{meV}$ and $-130 \, \mathrm{meV}$, respectively.
Intriguingly, the DFT+$U$ calculations predict a heavy Dirac point at the $M$ point that originates predominantly from the Ce-4$f$ orbitals.

\begin{figure}[tbp]
 \begin{center}
  \includegraphics[keepaspectratio, scale=0.37]{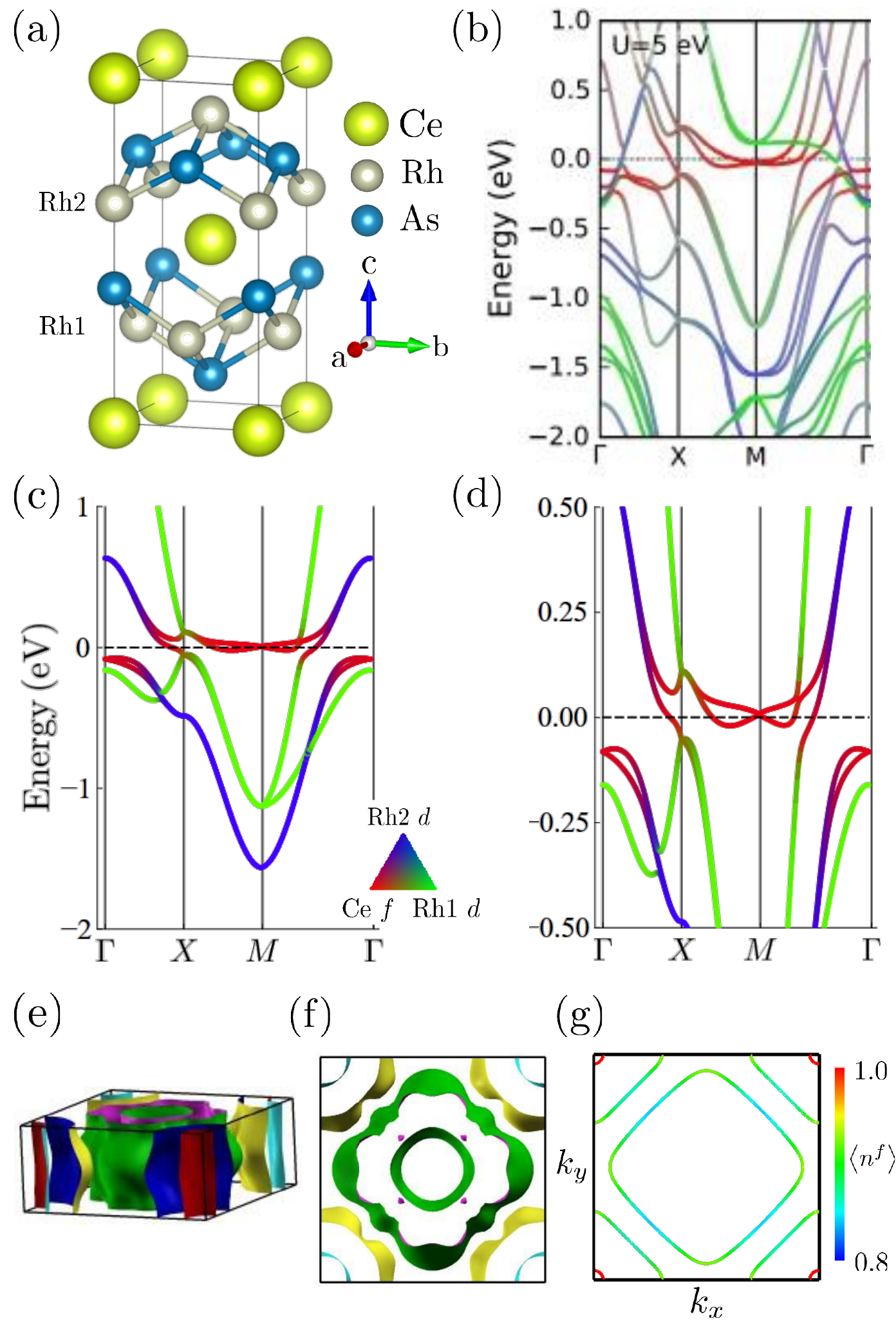}
  \end{center}
  \caption{
  (a) Crystalline structure of CeRh$_2$As$_2$. The Rh1 and Rh2 sites are indicated. The figure is adapted from Ref.~\cite{Nogaki2021prr}.
  (b) The DFT+$U$ band structure calculations for $U=5\,\mathrm{eV}$. The band structure along the $\Gamma$-$X$-$M$-$\Gamma$ line is shown. The orbital weight of Ce-4$f$, Rh1-5$d$, and Rh2-5$d$ orbitals are indicated by red, green, and blue, respectively. 
  (c) The band structure of the Dirac-Anderson model.
  (d) The enlarged view of the Dirac-Anderson model around the Fermi level, $E_{\mathrm{F}}$.
  (e) The Fermi surface obtained from the DFT+$U$ calculation for $U=5\,\mathrm{eV}$.
  (f) The top view of the Fermi surface from the DFT+$U$ calculation.
  (g) The Fermi surface of the Dirac-Anderson model. The color indicates the weight of $f$-orbitals.
  Figures~(b), (e) and (f) are adapted from Ref.~\cite{Ishizuka2024prb}.
  }
  \label{fig:band}
\end{figure}

To construct a minimal model of CeRh$_2$As$_2$, we took one of each of the Ce-4$f$, Rh1-5$d$, and Rh2-5$d$ orbitals and constructed a two-dimensional 12-orbital \textit{Dirac-Anderson model} that includes spin and sublattice degrees of freedom.
Figures~\ref{fig:band}(b-d) compare the band dispersions obtained from the DFT+$U$ calculation with band dispersions of the one-body part of the Dirac-Anderson model. 
The global and local band structures and the orbital character of each band of the model, particularly near the Fermi level~$E_{\mathrm{F}}$, are in approximate agreement with the DFT+$U$ calculation. 
Importantly, the van Hove singularity around the $X$ point and the heavy Dirac point at the $M$ point are both reproduced.
Because the experimentally observed van Hove singularities lie slightly above the DFT+$U$ result, we introduce renormalization factors $z = 0.30$ and $\tilde z = 0.3464$ into the $f$-orbital hopping and $c$-$f$ hybridization, respectively (see Appendix~\ref{sec:tight}).

Figures~\ref{fig:band}(e-g) show the shapes of the Fermi surfaces obtained from the DFT+$U$ calculation and from the Dirac-Anderson model. 
The large Fermi surface around the $\Gamma$ point consists mainly of conduction electrons, whereas the small Fermi surfaces around the $M$ point originate from the heavy Dirac band. 
Thus, bands with markedly different effective masses coexist in this system. 
The small cylindrical Fermi surface around the $\Gamma$ point is omitted from our model because it is expected to be not important for magnetism and superconductivity due to its weakly correlated and dispersive character. 
The detailed definition and the adapted parameters of the Dirac-Anderson model is provided in Appendix~\ref{sec:tight}.
The Hubbard-type on-site Coulomb repulsion between $f$-electrons is introduced.

The constructed model is analyzed using the random-phase approximation (RPA) and the fluctuation-exchange (FLEX) approximation.
For both numerical techniques, we use $1024\times1024$ $\bk$-mesh.
The temperature is set to $T=1\times10^{-4}$, comparable to the superconducting transition temperature of CeRh$_2$As$_2$, $T_{\rm c} \sim 0.3\,\mathrm{K}$. 
The cutoff parameter $\Lambda$ of the singular value decomposition is set to $1\times10^6$, satisfying the relation $\beta \omega_{\mathrm{max}} < \Lambda$.
Here, $\beta$ and $\omega_{\mathrm{max}}$ are the inverse temperature and the ultraviolet cutoff of the band structure.
All quantities with a dimension of energy are defined in units of electronvolts (eV).

\section{Multipole fluctuation}
\label{sec:suscep}
The degrees of freedom of the system can be systematically classified by the augmented multipole basis~\cite{Watanabe2018prb,Hayami2018prb,Yatsushiro2021prb}. 
Since the on-site Hubbard interaction $U$ is taken into account only for Ce-4$f$ electrons, the multipole operators built from Ce-4$f$ degrees of freedom describe fluctuations enhanced by electron-correlation effects. 
The multipole operators of Ce-4$f$ electrons are given by the tensor product,
\begin{align}
  \mathcal{Q}^{\mu\nu} = \bar{s}^{\mu} \otimes \bar{\sigma}^{\nu},
\end{align}
where $\bar{s}$ and $\bar{\sigma}$ represent the normalized Pauli and unit matrices in the spin and sublattice spaces, respectively~\cite{Nogaki2022prb,Nogaki2024prb}. 
Here, the normalization condition $\mathrm{tr}[\mathcal{Q}^\dagger \mathcal{Q}] = 1$ is imposed on the multipole operators. 
The classification of the multipole operators in our model is summarized in Table~\ref{tab:multi_op}. 
Because fluctuations of inter-sublattice multipoles are negligibly small, they are omitted from this classification table. 
By virtue of the sublattice degrees of freedom, odd-parity multipoles proportional to $\sigma_z$ could become active.
\begin{table}[tbp]
    \centering
    \begin{tabular}{cc} 
    \hline\hline
    operator & multipole \\ 
    \hline
    $s^0\otimes\sigma^0$ & Electric monopole \\ 
    $s^0\otimes\sigma^z$ & Electric dipole \\ 
    $s^z\otimes\sigma^0$ & Magnetic dipole \\ 
    $s^z\otimes\sigma^z$ & Magnetic monopole \\ 
    $s^{\pm}\otimes\sigma^0$ & Magnetic dipole \\ 
    $s^{\pm}\otimes\sigma^z$ & Magnetic quadrupole \\ 
    \hline \hline
    \end{tabular}
    \caption{
    Classification of the multipole operators in the Dirac-Anderson model.
    Here, $s^{\pm} = (s^x \pm i s^y)/\sqrt{2}$ represent the ladder operators in spin space.
    Although higher-order multipole bases are mixed in the crystalline environment, only the lowest-order multipole basis is shown.
    For example, $s^z \otimes \sigma^z$ is a linear combination of the magnetic monopole and magnetic quadrupole operators in the lattice described by the point group $D_{4h}$.
    }
    \label{tab:multi_op}
\end{table}

Multipole fluctuations are described by correlation functions.  
The correlation function for a multipole $\mathcal Q$ is defined as
\begin{align}
  \chi^{\mathcal Q}(q) = \sum_{\{\xi\}}\mathcal Q_{\xi_2\xi_1}\,
  \chi_{\xi_1\xi_2\xi_3\xi_4}(q)\,
  \mathcal Q_{\xi_3\xi_4},
\end{align}
where $\chi_{\xi_1\xi_2\xi_3\xi_4}$ is the generalized susceptibility tensor.  
Here, $q=(\bm q,i\nu_n)$ and $\xi=(s,\sigma)$ are abbreviated notation.
The $i\nu_n$ represent the bosonic Matsubara frequencies.

\begin{figure}[tbp]
 \begin{center}
\includegraphics[keepaspectratio, scale=0.41]{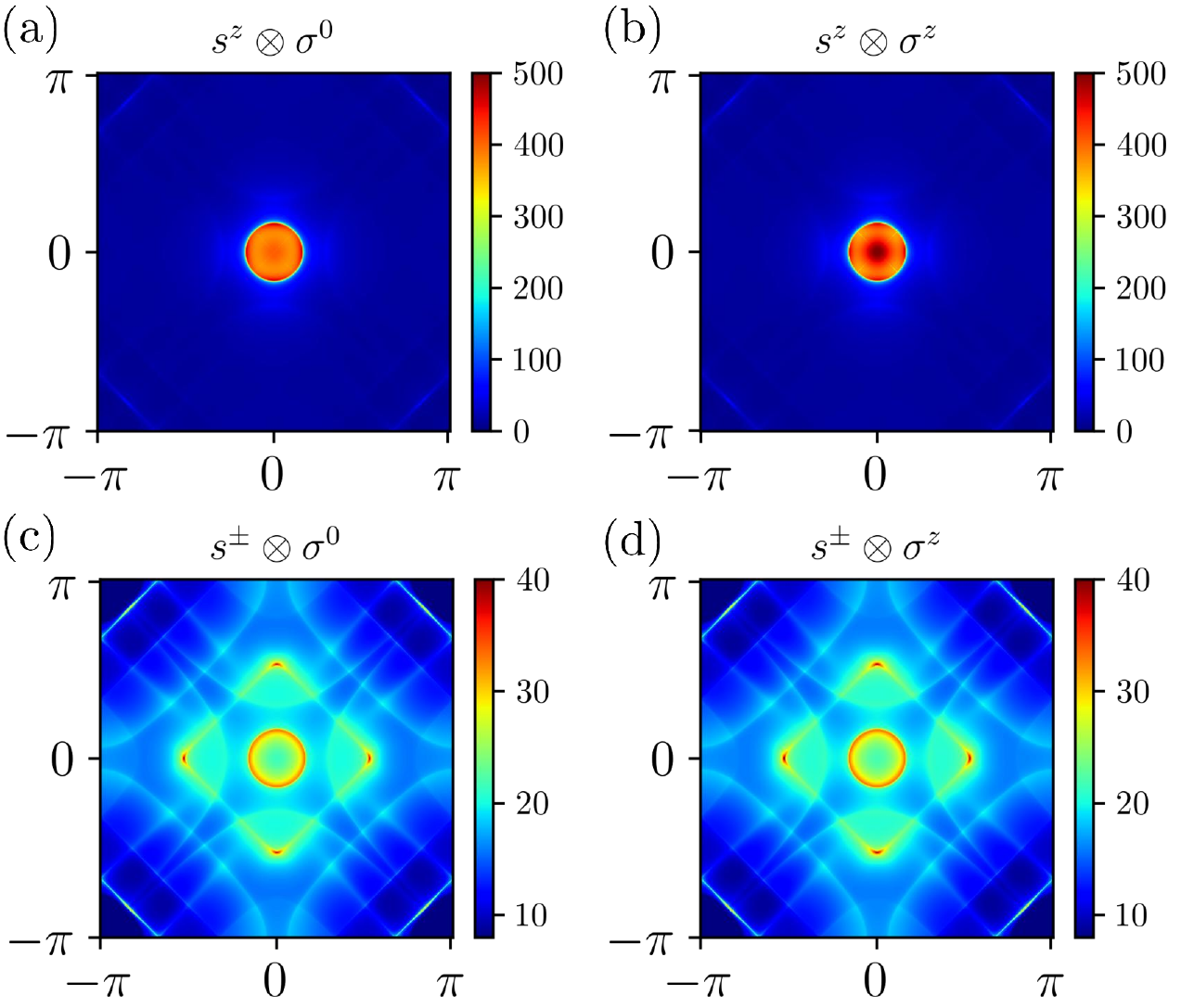}
  \end{center}
  \caption{
  The momentum dependence of the static multipole susceptibilities $\chi^{\mathcal Q}(\bm{q},i\nu_n=0)$ for the Stoner factor $\alpha = 0.985$.
  The definition of the Stoner factor is given in Sec.~\ref{sec:sc}.
  (a) Even-parity longitudinal magnetic multipole $s^z \otimes \sigma^0$.
  (b) Odd-parity longitudinal magnetic multipole $s^z \otimes \sigma^z$.
  (c) Even-parity transverse magnetic multipole $s^{\pm} \otimes \sigma^0$.
  (d) Odd-parity transverse magnetic multipole $s^{\pm} \otimes \sigma^z$.
  The color scale of panel (a) [panel (c)] is the same as panel (b) [panel (d)].
  }
  \label{fig:suscep}
\end{figure}

In Fig.~\ref{fig:suscep}, we present the momentum dependence of the representative multipole susceptibilities evaluated within the RPA.  
Both even- and odd-parity longitudinal magnetic fluctuations [Figs.~\ref{fig:suscep}(a) and \ref{fig:suscep}(b)] show a pronounced peak around the $\Gamma$ point, indicating ferroic behavior.  
Because the odd-parity longitudinal fluctuation, classified as the magnetic monopole fluctuation, has a larger susceptibility than the even-parity one, ferroic magnetic monopole fluctuations are dominant in the Dirac-Anderson model.
The transverse susceptibilities [Figs.~\ref{fig:suscep}(c) and \ref{fig:suscep}(d)] are roughly an order of magnitude smaller than the longitudinal ones.  
Their momentum dependence is more intricate: Ring- and square-shaped structures appear around the $\Gamma$ point, together with a nearly antiferroic peak at $(\pi-\delta,\pi-\delta)$.  
Overall, the multipole fluctuations in our model are governed by ferroic magnetic monopole fluctuations.  
Since the ferroic magnetic monopole order has been observed inside the superconducting phase in the NMR/NQR measurements~\cite{Kibune2022prl,Ogata2024prb}, our results are consistent with the experiment.

Let us discuss the origin of the magnetic monopole fluctuations.  
As emphasized in Sec.~\ref{sec:model}, the Dirac point originating from Ce-4$f$ electrons is present at the $M$ point.  
The Dirac point near the Fermi level $E_{\mathrm F}$ enriches the geometric properties of the system: the associated Dirac band contributes largely to the quantum metric.  
As pointed out in Refs.~\cite{Kitamura2024prl,Kudo2025arxiv}, the sizable quantum metric suppresses finite $\bm{q}$ fluctuations, resulting in ferroic behaviors.  
Although an analytic expression is available only for the SU(2)-symmetric cases~\cite{Kitamura2024prl,Kudo2025arxiv}, quantum geometry should play a pivotal role in the spin-orbit coupled Dirac-Anderson model as well.  
Therefore, it is expected that magnetic monopole fluctuations are stabilized by quantum geometry.

\section{superconductivity}
\label{sec:sc}
Utilizing the linearized Éliashberg equation, we investigate the superconducting state:
\begin{align}
  \lambda \Delta_{\xi \xi'}(k) &= \frac{T}{N} \sum_{k'} \Gamma_{\xi \xi_1 \xi_2 \xi'}^a\left(k-k'\right) F_{\xi_1 \xi_2}(k'), \\
  F_{\xi_1 \xi_2}(k) &= -G_{\xi_1 \xi_3}\left(k\right) \Delta_{\xi_3 \xi_4}\left(k\right) G_{\xi_2 \xi_4}\left(-k\right),
\end{align}
where the abbreviated notation $k=(\bm{k},i\omega_n)$ is adapted.
Here, $\lambda$ is the eigenvalue of the linearized \'{E}liashberg equation,  
$\Delta(k)$ is the superconducting order parameter, $\Gamma^{a}$ is the irreducible particle-particle vertex, $G(k)$ is the single-particle Green function, $F(k)$ is the linearized anomalous Green function, and $i\omega_n$ is the fermionic Matsubara frequency.  
The superconducting instability occurs when $\lambda = 1$.
\begin{figure}[tbp]
 \begin{center}
\includegraphics[keepaspectratio, scale=0.41]{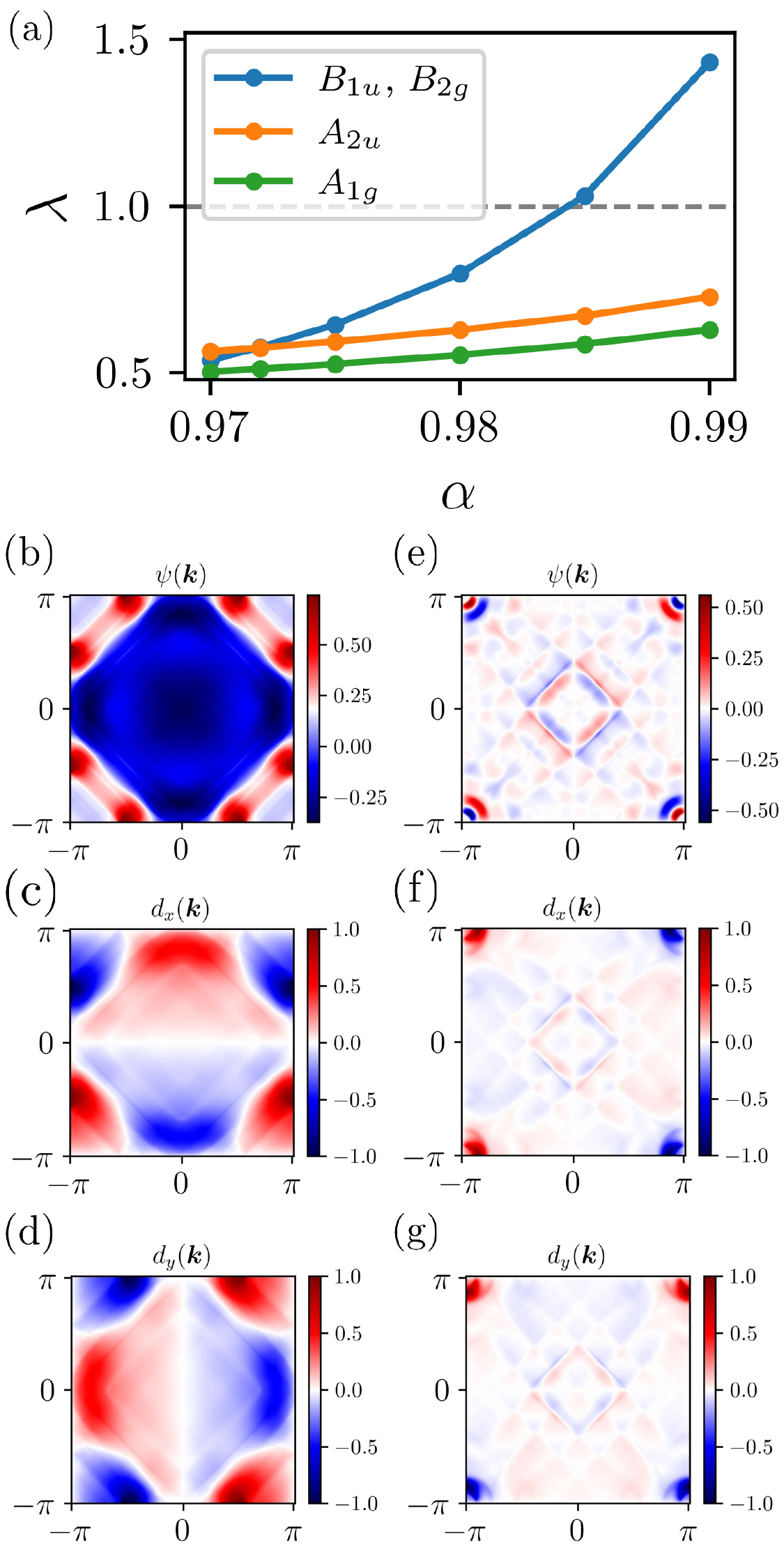}
  \end{center}
  \caption{
  (a) Eigenvalues $\lambda$ of the \'{E}liashberg equation as a function of the Stoner factor $\alpha$. The eigenvalue for the $B_{2g}$ representation is nearly degenerate with the eigenvalue of the inversion partner, $B_{1u}$.
  (b-d) Spin-singlet component $\psi(\bm k)$ and spin-triplet components $\bm d(\bm k)$
         of the gap function in the $A_{2u}$ state.  
  (e-g) Corresponding components in the $B_{1u}$ state.
  Note that $z$-component of the spin-triplet component $d_z(\bk)$ is prohibited by the mirror symmetry $\sigma_h$ in the two-dimensional system.
  }
  \label{fig:delta}
\end{figure}

The leading superconducting instabilities belong to the $A_{1g}$, $A_{2u}$, $B_{2g}$, and $B_{1u}$ irreducible representations of the point group $D_{4h}$.  
The inversion symmetry partners of $B_{2g}$/$B_{1u}$ possess almost degenerate eigenvalues, whereas others of $A_{1g}$/$A_{2u}$ have eigenvalues whose degeneracy is lifted but remain close.
The following relations $\lambda_{B_{1u}} > \lambda_{B_{2g}}$ and $\lambda_{A_{2u}} > \lambda_{A_{1g}}$ hold for all values of the Coulomb interaction $U$ considered in this paper, indicating that the odd-parity superconducting states are stable.
The leading state changes as the Hubbard interaction $U$ is varied.
Figure~\ref{fig:delta}(a) plots $\lambda$ versus the Stoner factor defined as,
\begin{align}
  \alpha = \max_q D\!\bigl[\hat\chi^0(q)\,\hat U\bigr],
\end{align}
where $\hat\chi^{0}(q)$ is the bare susceptibility matrix, $\hat U$ is the interaction matrix, and $D[A]$ returns the largest eigenvalue of~$A$.  
The Stoner factor~$\alpha$ increases monotonically with~$U$; $\alpha=1$ signals a multipole phase transition.  
While the $A_{1g}$/$A_{2u}$ states are dominant for small~$U$, the $B_{2g}$/$B_{1u}$ eigenvalues grow rapidly and become leading for large~$U$.  
At $\alpha\simeq0.985$, $\lambda_{B_{1u}}$ reaches unity, indicating a superconducting transition in the $B_{1u}$ channel.

Figures~\ref{fig:delta}(b)-(g) show the momentum dependence of the gap functions in the $B_{1u}$ and $A_{2u}$ states,
\begin{equation}
  \Delta(\bm k)=
    \psi(\bm k)\,is_y\!\otimes\!\sigma_z
  + \bm d(\bm k)\!\cdot\!\bm s\,is_y\!\otimes\!\sigma_0,
\end{equation}
where $\psi(\bm k)$ is a sublattice-antisymmetric spin-singlet component and
$\bm d(\bm k)$ is a sublattice-symmetric spin-triplet component.  
In both $A_{2u}$ and $B_{1u}$ states, the spin-triplet component dominates, so these states are essentially spin-triplet pairing states, mixed with a minor but comparable spin-singlet pairing due to local inversion symmetry breaking at Ce sites.
The spin-triplet character is a direct consequence of the ferroic magnetic fluctuations discussed in Sec.~\ref{sec:suscep}.  
Whereas the gap function of the $A_{2u}$ state has a large amplitude on the large Fermi surface, especially near the van Hove singularities on the $\Gamma$-$X$ line, the gap function of the $B_{1u}$ state is concentrated on the small Fermi surface originating from the Dirac point at the $M$~point.
Hence, the Dirac point plays an essential role in both the magnetic fluctuations and the superconductivity through quantum geometry and the large density of states.

References~\cite{Cavanagh2022prb,Suh2023prr,Lee2025prl} have pointed out that the spin-orbit coupling dominates over the inter-sublattice hopping at the Brillouin-zone edge due to the nonymmorphic crystalline structure.
The gap functions of the $B_{2g}/B_{1u}$ states are strongly localized at the Brillouin-zone edge, signaling that the inter-sublattice hopping is irrelevant to the thermodynamic stability of the $B_{2g}/B_{1u}$ states.
Since the inter-sublattice hopping lifts the degeneracy of even- and odd-parity superconducting states, the $B_{2g}/B_{1u}$ states should possess nearly degenerate eigenvalues, which is consistent with Fig.~\ref{fig:delta}(a).
By contrast, gap functions of the $A_{1g}/A_{2u}$ states show large amplitudes away from the Brillouin-zone edge and therefore the degeneracy of their eigenvalues is lifted.

In CeRh$_2$As$_2$, a field-induced phase transition from even-parity to odd-parity superconductivity is observed~\cite{Khim2021science}: the order parameter is even-parity in low fields while odd-parity in high fields.  
At $\alpha=0.985$ the $B_{1u}$ and $B_{2g}$ states have nearly degenerate eigenvalues, $\lambda_{B_{1u}}=1.0314$ and $\lambda_{B_{2g}}=1.0306$.  
The approximate degeneracy is consistent with CeRh$_2$As$_2$, but there seems to be a contradiction because in the Dirac-Anderson model the odd-parity superconductivity can be stabilized without a magnetic field. This discrepancy could be resolved by the renormalization effect discussed in the next section.

\section{Effect of Coulomb interaction on quantum geometry}
\label{sec:flex}
\begin{figure}[tbp]
  \centering
  \includegraphics[keepaspectratio,scale=0.35]{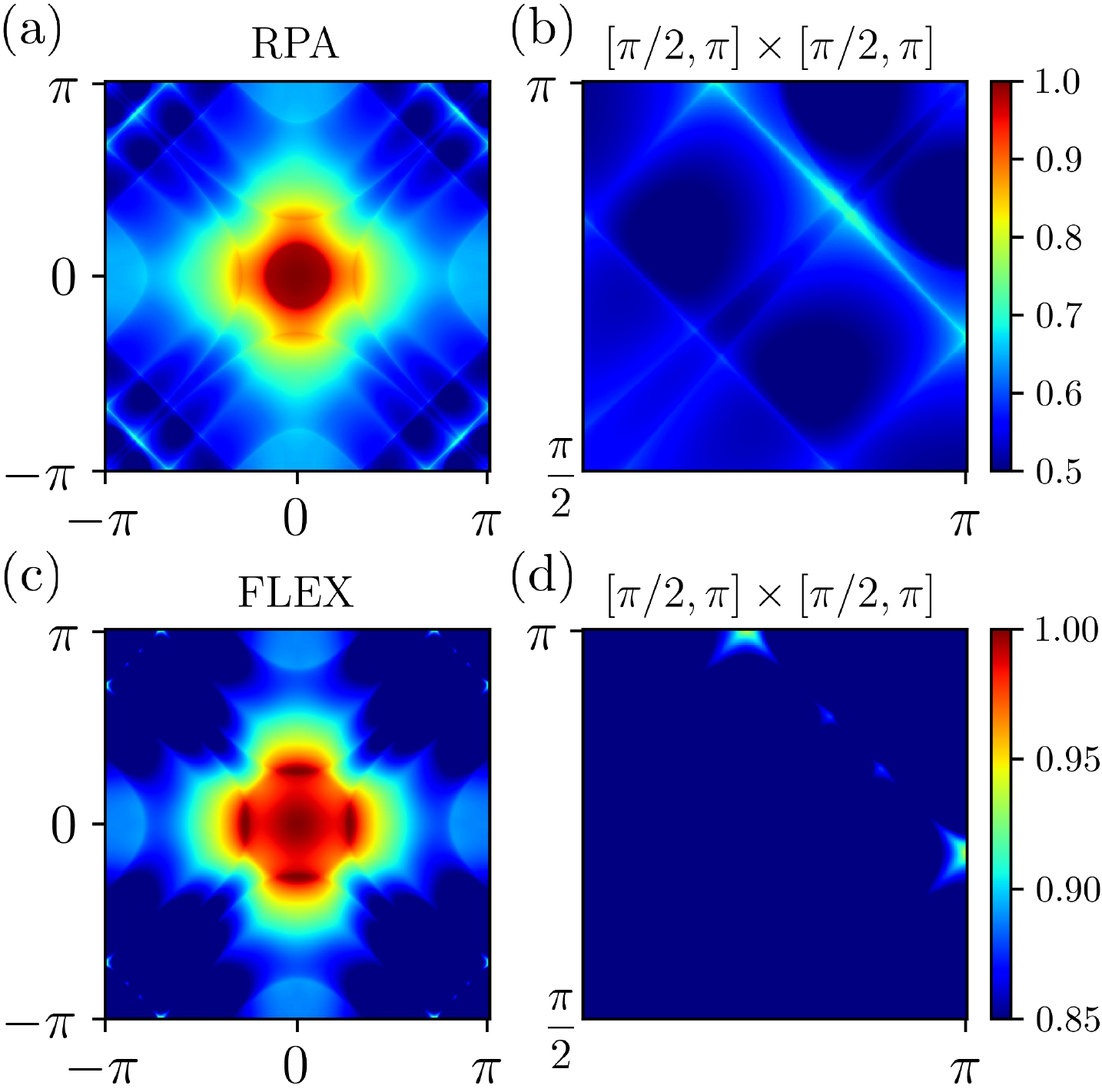}
  \caption{%
  Comparison between the RPA and FLEX.
  Normalized momentum dependence of the bare susceptibility $\chi^{0}(\bm q)/\max_{\bm{q}}\chi^{0}(\bm q)$ for the magnetic monopole operator $s^{z}\!\otimes\!\sigma^{z}$ is shown. Each dataset is normalized to the maximum value $\max_{\bm{q}}\chi^{0}(\bm q)$.  
  (a) Susceptibility calculated within the RPA.  
  (b) The enlarged view of (a) around $[\pi/2, \pi]\times[\pi/2, \pi]$.
  (c) Susceptibility calculated by the FLEX approximation for $U=0.22$.
  (d) The enlarged view of (c) around $[\pi/2, \pi]\times[\pi/2, \pi]$.
  The color scale of panel (a) [panel (c)] is the same as panel (b) [panel (d)].
  }
  \label{fig:flex}
\end{figure}

In this section, we discuss the renormalization effect on susceptibility that is significantly affected by quantum geometry.
The FLEX approximation, the self-consistent extension of the RPA, includes the self-energy correction, and therefore, comparison between these two calculations should be a touchstone of robustness of the quantum geometric effect against the strong-correlation effect.
In Fig.~\ref{fig:flex}, the normalized bare susceptibilities 
$\chi^0(\bm{q})/\max_{\bm{q}}\chi^0(\bm{q})$ of the magnetic monopole operator $s^z\otimes\sigma^z$ calculated by the RPA and FLEX approximation are compared.
In the FLEX approximation, the Hubbard interaction $U=0.22$ is adopted where the Stoner factor is $\alpha \simeq 0.995$.
The overall momentum dependence of these susceptibilities is qualitatively similar; in particular, the large intensity at the $\Gamma$ point remains despite the self-energy correction.
For quantitative comparison, we compare the fluctuation around $M$ point $\chi^0(\bm{q} \sim (\pi-\delta,\pi))$ calculated by the two techniques as follows:
\begin{align}
  \frac{\chi^0(\bm{q}\sim(\pi-\delta,\pi))}{\max_{\bm{q}}\chi^0(\bm{q})}
  \sim
  \left\{
  \begin{array}{ll}
  0.72 & (\mathrm{RPA}) \\
  0.94 & (\mathrm{FLEX})
  \end{array}
  \right.,
\end{align}
indicating that antiferroic fluctuations are recovered as a result of the self-energy correction.
Since the effect of the quantum metric suppresses finite $\bm{q}$ fluctuations~\cite{Kitamura2024prl}, this result indicates that the renormalization effect due to the Coulomb interaction suppresses the effect of quantum geometry.
This is qualitatively consistent with Ref.~\cite{Sukhachov2025prb}, where the suppression of the quantum metric for many-body systems is demonstrated in perfectly and nearly flat-band models.
Thus, we conclude that the Coulomb interaction suppresses the effect of quantum geometry and moderates the ferroic character of the magnetic susceptibility.

The results of the FLEX approximation indicate that ferroic magnetic-monopole and antiferromagnetic fluctuations coexist as a result of the mitigation of quantum geometric effects.
The neutron scattering measurements for CeRh$_2$As$_2$ also indicate the presence of strong antiferromagnetic fluctuations~\cite{Chen2024prl}.  
Therefore, by including strong-correlation effects, the Dirac-Anderson model could agree with the observations by the neutron scattering experiments.

The Coulomb interaction also changes the magnetic anisotropy discussed in Sec.~\ref{sec:suscep}. 
While the magnetic anisotropy calculated by the RPA is estimated as  $\max_{\bm{q}}\chi^0_{s^z\otimes\sigma^z}(\bm{q})/ \max_{\bm{q}}\chi^0_{s^\pm\otimes\sigma^z}(\bm{q}) \sim 1.33$, the FLEX approximation yields smaller anisotropy $\max_{\bm{q}} \chi^0_{s^z\otimes\sigma^z}(\bm{q})/\max_{\bm{q}} \chi^0_{s^\pm\otimes\sigma^z}(\bm{q}) \sim 1.06$.
Here, $\chi^0_{s^z\otimes\sigma^z}(\bm{q})$ and $\chi^0_{s^\pm\otimes\sigma^z}(\bm{q})$ represent the bare susceptibilities of the magnetic monopole and magnetic quadrupole operator, respectively.

Since the dressed magnetic fluctuations show a relatively large antiferromagnetic intensity, the spin-singlet channel would be favored compared to the RPA results.  
Furthermore, suppression of the magnetic anisotropy could also favor the spin-singlet channel~\cite{Yanase2003pr}.
These changes will increase $\lambda_{B_{2g}}$ relative to $\lambda_{B_{1u}}$, making the even-parity state the zero-field superconducting phase, consistent with CeRh$_2$As$_2$ at ambient pressure.
In the experimental results, applying the external pressure in CeRh$_2$As$_2$ brings the transition temperatures of even- and odd-parity superconductivity close to each other~\cite{Siddiquee2023prb,Semeniuk2023prb}.
Applied pressure is expected to weaken the renormalization effect and make the RPA result more reliable.
Thus, the Dirac-Anderson model analyzed by the RPA likely provides a more precise description of high-pressured CeRh$_2$As$_2$, where the even- and odd-parity states are almost degenerate. They are identified with the $B_{2g}$ and $B_{1u}$ ($d_{xy}+p$-wave) representations, respectively.

\section{Conclusion}
\label{sec:summary}
We have constructed a two-dimensional 12-orbital Dirac-Anderson model that reproduces the ARPES spectra and DFT+$U$ band structure of CeRh$_2$As$_2$, including the van Hove singularities and the $M$-point $f$-electron Dirac point.  
The quantum geometry of the Dirac point favors the ferroic magnetic monopole fluctuation.
Linearized Eliashberg calculations on this model reveal four nearly degenerate superconducting instabilities—$A_{1g}$, $A_{2u}$, $B_{2g}$, and $B_{1u}$. 
As the Stoner factor approaches the multipole-quantum-critical value, the spin-triplet dominant odd-parity $B_{1u}$ and even-parity $B_{2g}$ states become dominant.

A comparison between the RPA and FLEX approximation shows that strong-correlation effects renormalize the quantum-geometric suppression of finite $\bm{q}$ fluctuations and the magnetic anisotropy, producing a coexistence of dominant ferroic magnetic monopole and subdominant antiferromagnetic fluctuations that are consistent with NMR/NQR and neutron scattering experiments.  
Because hydrostatic pressure tends to suppress the renormalization effect, our results provide a natural explanation for the pressure-induced even-odd parity transition suggested in CeRh$_2$As$_2$~\cite{Siddiquee2023prb,Semeniuk2023prb}.

\begin{acknowledgments}
K.N. was supported by JSPS KAKENHI Grant No.~JP24K22869.
Y.Y. was supported by JSPS KAKENHI Grant Nos.~JP23K22452, JP22H04933, JP23K17353, JP24K21530, JP24H00007, and JP25H01249.
We appreciate helpful discussions with Jun Ishizuka, Taisei Kitamura, and Elena Hassinger.
The numerical calculations were carried out on Yukawa-21 at YITP in Kyoto University.
Some figures are generated by \texttt{VESTA}~\cite{Momma2011jac}, \texttt{XCrySDen}~\cite{Kokaj1999jmgm}, and \texttt{FermiSurfer}~\cite{Kawamura2019cpc}.
\end{acknowledgments}

\appendix

\section{Details of the tight-biding model}
\label{sec:tight}
The Dirac-Anderson model consists of the one-body term and the two-body term:
\begin{align}
  H = H_{0} + H_{U}.
\end{align}
The one-body Hamiltonian is decomposed into Ce-4$f$ Hamiltonian, Rh1-5$d$ Hamiltonian, Rh2-5$d$ Hamiltonian, and the hybridization term: 
\begin{align}
  H_0 &= \sum_{\bk,\eta,\eta'} c^\dagger_{\bk,\eta} \mathcal{H}_{\eta,\eta'}(\bk) c_{\bk,\eta'}, \\
  \mathcal{H}(\bk) &=
	\begin{pmatrix}
    \mathcal{H}^{f}_{\bk} & \mathcal{H}^{f-d1}_{\bk} & \mathcal{H}^{f-d2}_{\bk} \\
    \mathcal{H}^{f-d1,\dagger}_{\bk} & \mathcal{H}^{d1}_{\bk} & 0 \\
    \mathcal{H}^{f-d2,\dagger}_{\bk} & 0 & \mathcal{H}^{d2}_{\bk}
  \end{pmatrix},
\end{align}
where $c^\dagger_{\bk,\eta}$ creates the electron with the momentum $\bm{k}$ and index $\eta = (\tau,s,\sigma)$ of the orbital $\tau$, spin $s$, and sublattice $\sigma$.
Each term of the one-body Hamiltonian is given by:
\begin{align}
  \mathcal{H}^{f}_{\bk} &= \varepsilon^{f}_{\bk} s_0 \otimes \sigma_0 + \bm{g}^{f}_{\bk}\cdot\bm{s} \otimes \sigma_z \notag \\
  &+ t^{f}_{\perp,\bk} s_0 \otimes \sigma_{+} + t^{f}_{\perp,-\bk} s_0 \otimes \sigma_{-}, \\
  \mathcal{H}^{d1}_{\bk} &= \varepsilon^{d1}_{\bk} s_0 \otimes \sigma_0\notag \\
  &+ t^{d1}_{\perp,\bk} s_0 \otimes \sigma_{+} + t^{d1}_{\perp,-\bk}  s_0 \otimes \sigma_{-}, \\
  \mathcal{H}^{d2}_{\bk} &= \varepsilon^{d2}_{\bk} s_0 \otimes \sigma_0, \\
  \mathcal{H}^{f-d1}_{\bk} &= s_0 \otimes 
	\begin{pmatrix}
  \tilde{t}^{f-d1}_{\perp,k_x} & \tilde{t}^{f-d1}_{\perp,k_y} \\
  \tilde{t}^{f-d1}_{\perp,-k_y} & \tilde{t}^{f-d1}_{\perp,-k_x}
	\end{pmatrix}  \notag \\
  &+ s_x \otimes 
	\begin{pmatrix}
  0 & \tilde{g}^{f-d1}_{k_y} \\
  -\tilde{g}^{f-d1}_{-k_y} & 0
	\end{pmatrix}  \notag \\
  &-s_y\otimes 
	\begin{pmatrix}
  \tilde{g}^{f-d1}_{k_x} & 0 \\
  0 & -\tilde{g}^{f-d1}_{-k_x}
	\end{pmatrix} , \\
  \mathcal{H}^{f-d2}_{\bk} &= t^{f-d2}_{\perp,\bk} s_0 \otimes \sigma_{+} + t^{f-d2}_{\perp,-\bk} s_0 \otimes \sigma_{-}.
\end{align}
Here, intra-sublattice hopping ($\varepsilon$), inter-sublattice hopping ($t_{\perp}$, $\tilde{t}_{\perp}$), and spin-orbit coupling ($\bm{g}$, $\tilde{g}$) are given by: 
\begin{align}
  \varepsilon_{\bk} &= 2t_1(\cos k_x+\cos k_y) + 4t_2\cos k_x\cos k_y \notag \\
  &+ 2t_3(\cos 2k_x+\cos 2k_y) + \mu, \\
  t_{\perp,\bk} &= t_{\perp,1}(1+e^{-ik_x})(1+e^{-ik_y}) \notag \\ 
  &+ t_{\perp,2}\left\{(e^{-2ik_y}+e^{ik_y})(1+e^{-ik_x}) + (k_x \leftrightarrow k_y)\right\}, \\
  g_{x,\bk} &= 2\alpha\sin k_y, \\
  g_{y,\bk} &= -2\alpha\sin k_x, \\
  \tilde{t}_{\perp,k} &= \tilde{t}_{\perp}(1+e^{-ik}), \\
  \tilde{g}_{k} &= \tilde{\alpha}i(-1 + e^{-ik}).
\end{align}
The hopping parameters are summerized in Tab.~\ref{tab:tight}.
The filling in the unit cell is fixed to 7.5.
To take into account the mass renormalization, we employ the renormalization factor $z=0.3$ and $\tilde{z} = 0.3464$.
The hopping parameter between $f$-orbitals and the hybridization parameter are renormalized by $z$ and $\tilde{z}$, respectively.

The two-body term includes the on-site Hubbard-type interaction between $f$-orbitals:
\begin{align}
  H_U = U \sum_{i,\sigma} n^f_{i\uparrow\sigma}n^f_{i\downarrow\sigma},
\end{align}
where $n^f_{i s \sigma} = c^\dagger_{i f s \sigma} c_{i f s \sigma}$ represents the $f$-electron density operator of the site $i$, the spin $s$, and the sublattice $\sigma$.

\begin{table}[tbp]
    \centering
    \begin{tabular}{cccccc}
    \hline\hline
     & $f$ & $d1$ & $d2$ & $f$-$d1$ & $f$-$d2$ \\ 
    \hline
    $\mu$ & -0.085 & 0.05 & -0.475 & & \\ 
    $t_1$ & -0.016 & 0.29 & 0.27 & & \\ 
    $t_2$ & -0.016 & & & & \\ 
    $t_3$ & 0.012 & & & & \\ 
    $\alpha$ & -0.05 & & & & \\ 
    $t_{\perp,1}$ & & 0.34 & & & 0.07 \\ 
    $t_{\perp,2}$ & & & & & -0.04 \\ 
    $\tilde{\alpha}$ & & & & 0.1 & \\ 
    $\tilde{t}_{\perp}$ & & & & 0.06 & \\ 
    \hline \hline
    \end{tabular}
    \caption{The list of the tight-biding parameters. The blank element means 0.}
    \label{tab:tight}
\end{table}

\begin{figure}[tbp]
 \begin{center}
\includegraphics[keepaspectratio, scale=0.47]{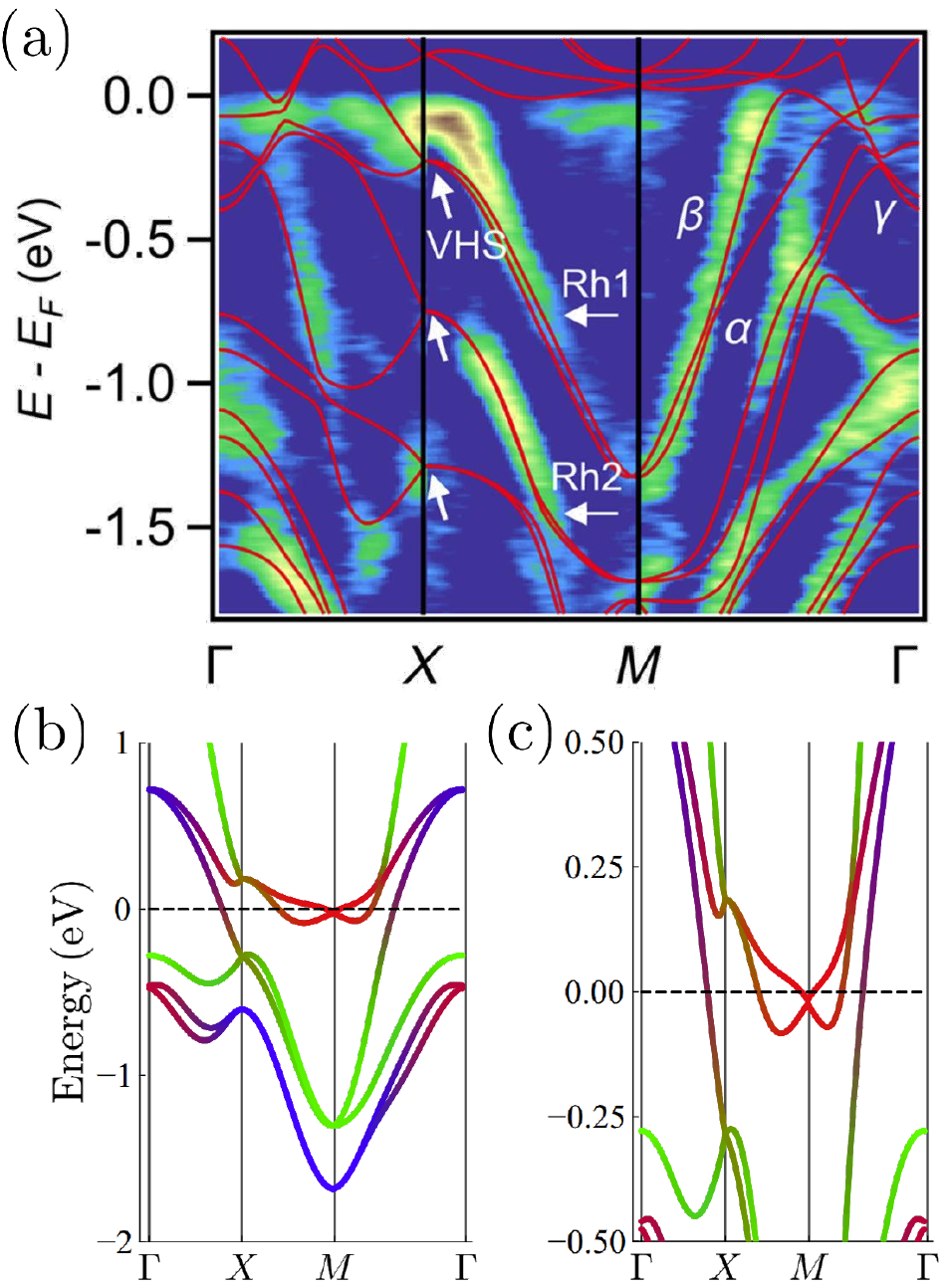}
  \end{center}
  \caption{
  (a) The ARPES data (intensity) with DFT+$U$ calculation (red line). The figure is adapted from Ref.~\cite{Chen2024prx}.
  (b) The band structure of the Dirac-Anderson model without renormalization factor $z$ and $\tilde{z}$.
  (c) The enlarged view of (b).
  }
  \label{fig:app_band}
\end{figure}

\section{Random-phase approximation and fluctuation exchange approximation}
\label{sec:rpa_flex}
The noninteracting Green functions for $U=0$ are expressed
by matrix form in the orbital, spin, and sublattice basis,
\begin{equation}
  \label{eq:noint_green}
  G^{(0)}(\bm{k},i\omega_n) = \left(i\omega_n I - \mathcal{H}_{\bk}\right)^{-1},
\end{equation}
where $\omega_n=(2n+1)\pi T$ and $\mathcal{H}_{\bk}$ are fermionic Matsubara frequencies and the one-body Hamiltonian.
Here, $I$ represents the unit matrix.
In the interacting case $U\neq0$, the dressed Green functions contain a self-energy $\Sigma(k)$, 
\begin{align}
  \label{eq:green_function_dressed}
  G(k) & = \left(i\omega_n I - \mathcal{H}_{\bk} -\Sigma(k)\right)^{-1}.
\end{align}
where the abbreviated expression $k=(\bm{k},i\omega_n)$ is adopted.
In the FLEX approximation, the self-energy is expressed with the use of an effective interaction, $\Gamma^n(q)$, as
\begin{align}
  \label{eq:self_energy}
  &\Sigma_{\xi\xi'}  (k) = \frac{T}{N} \sum_{q}
  \Gamma^n_{\xi\xi_1\xi'\xi_2}(q)G_{\xi_1\xi_2}(k-q),
\end{align}
and the effective interaction is given by
\begin{align}
  \label{eq:effective_interaction}
  \Gamma^n_{\xi_1\xi_2\xi_3\xi_4}(q) &= U_{\xi_1\xi_2\xi_5\xi_6} \chi_{\xi_5\xi_6\xi_7\xi_8}(q) U_{\xi_7\xi_8\xi_3\xi_4} \notag \\
  &-\frac{1}{2}U_{\xi_1\xi_2\xi_5\xi_6}\chi^{(0)}_{\xi_5\xi_6\xi_7\xi_8}(q)U_{\xi_7\xi_8\xi_3\xi_4},
\end{align}
where $U_{\xi_1\xi_2\xi_3\xi_4}$ is the bare interaction tensor that satisfies the following relation
\begin{align}
    \sum_{\xi_1\xi_2\xi_3\xi_4} U_{\xi_1\xi_2\xi_3\xi_4} c^\dagger_{\xi_1} c_{\xi_2} c_{\xi_3} c^\dagger_{\xi_4} = U \sum_{i,\sigma} n_{i\uparrow \sigma}n_{i\downarrow \sigma}, \\
  U_{\xi_1\xi_2\xi_3\xi_4} = \delta_{\sigma_1,\sigma_2}\delta_{\sigma_2,\sigma_3}\delta_{\sigma_3,\sigma_4} U_{s_1 s_2 s_3 s_4}, \\
  U_{\uparrow\downarrow\uparrow\downarrow} = U_{\downarrow\uparrow\downarrow\uparrow} = -U_{\uparrow\uparrow\downarrow\downarrow} = -U_{\downarrow\downarrow\uparrow\uparrow} = U.
\end{align}
Here, $\chi(q)$ is the generalized susceptibility.
With the bosonic Matsubara frequencies $i\nu_n$, the abbreviated notation $\xi=(s,\sigma)$ and $q=({\bm q}, i\nu_n)$ are employed. 
We introduce the bare susceptibility
\begin{align}
  \label{eq:bare_suscep}
  \chi^{(0)} _{\xi_1\xi_2\xi_3\xi_4}(q)= -\frac{T}{N}\sum_{k}G_{\xi_1\xi_3}(k) G_{\xi_4\xi_2}(k-q),
\end{align}
and compute the generalized susceptibility by
\begin{align}
  \label{eq:gener_suscep2}
  \chi_{\xi_1\xi_2\xi_3\xi_4}(q)&=\chi^{(0)}_{\xi_1\xi_2\xi_3\xi_4}(q) \notag \\
  &+\chi^{(0)}_{\xi_1\xi_2\xi_5\xi_6}(q)U_{\xi_5\xi_6\xi_7\xi_8}\chi_{\xi_7\xi_8\xi_3\xi_4}(q).
\end{align}
According to Eqs.~(\ref{eq:green_function_dressed})-(\ref{eq:gener_suscep2}),
$G$, $\Sigma$, $\Gamma^n$, $\chi^{(0)}$, and $\chi$
depend on each other, and therefore, we self-consistently determine these functions. The FLEX approximation is a conserving approximation in which several conservation laws are satisfied in the framework of the Luttinger-Ward theory~\cite{Luttinger1960pr,Luttinger1960pr2,Baym1961pr,Baym1962pr}.

The RPA approximation ignores the self-energy correction included in the gresed Green function in Eq.~(\ref{eq:green_function_dressed}).
In the bare susceptibilities in Eq.~(\ref{eq:bare_suscep}), the dressed Green function is replaced by the noninteracting Green function of Eq.~(\ref{eq:noint_green}).

For functions with fermionic Matsubara frequencies $A(\bm{q},i \omega_n)$, the static limit 
$A(\bm{q},0)$ is evaluated by an approximation justified at low temperatures,
\begin{equation}
  A(\bm{q},0) \simeq \frac{A(\bm{q},i\pi T)+A(\bm{q},-i\pi T)}{2}.
\end{equation}

For the analysis of the superconducting phase transition, the irreducible vertex function in the particle-particle channel $\Gamma^a$ is needed, and it is 
obtained by
\begin{align}
  \Gamma^a_{\xi_1\xi_2\xi_3\xi_4}(q) &= -\frac{1}{2} U_{\xi_1\xi_2\xi_3\xi_4} \notag \\
    &-U_{\xi_1\xi_2\xi_5\xi_6}\chi_{\xi_5\xi_6\xi_7\xi_8}(q)U_{\xi_7\xi_8\xi_3\xi_4}.
  \label{eq:pp_int}
\end{align}

For the matrix representations of $\hat{\chi}^{0}$ and $\hat{U}$, we regroup the four indices $\xi_{1},\xi_{2},\xi_{3},\xi_{4}$ into the pairs $(\xi_{1},\xi_{2})$ and $(\xi_{3},\xi_{4})$.
The two-index combinations $(\xi_{1},\xi_{2})$ are then arranged in row-major order: $(1,1),(1,2),\dots(1,n),(2,1),\dots(n,n)$.

\bibliography{paper}

\begin{thebibliography}{97}%
\makeatletter
\providecommand \@ifxundefined [1]{%
 \@ifx{#1\undefined}
}%
\providecommand \@ifnum [1]{%
 \ifnum #1\expandafter \@firstoftwo
 \else \expandafter \@secondoftwo
 \fi
}%
\providecommand \@ifx [1]{%
 \ifx #1\expandafter \@firstoftwo
 \else \expandafter \@secondoftwo
 \fi
}%
\providecommand \natexlab [1]{#1}%
\providecommand \enquote  [1]{``#1''}%
\providecommand \bibnamefont  [1]{#1}%
\providecommand \bibfnamefont [1]{#1}%
\providecommand \citenamefont [1]{#1}%
\providecommand \href@noop [0]{\@secondoftwo}%
\providecommand \href [0]{\begingroup \@sanitize@url \@href}%
\providecommand \@href[1]{\@@startlink{#1}\@@href}%
\providecommand \@@href[1]{\endgroup#1\@@endlink}%
\providecommand \@sanitize@url [0]{\catcode `\\12\catcode `\$12\catcode `\&12\catcode `\#12\catcode `\^12\catcode `\_12\catcode `\%12\relax}%
\providecommand \@@startlink[1]{}%
\providecommand \@@endlink[0]{}%
\providecommand \url  [0]{\begingroup\@sanitize@url \@url }%
\providecommand \@url [1]{\endgroup\@href {#1}{\urlprefix }}%
\providecommand \urlprefix  [0]{URL }%
\providecommand \Eprint [0]{\href }%
\providecommand \doibase [0]{https://doi.org/}%
\providecommand \selectlanguage [0]{\@gobble}%
\providecommand \bibinfo  [0]{\@secondoftwo}%
\providecommand \bibfield  [0]{\@secondoftwo}%
\providecommand \translation [1]{[#1]}%
\providecommand \BibitemOpen [0]{}%
\providecommand \bibitemStop [0]{}%
\providecommand \bibitemNoStop [0]{.\EOS\space}%
\providecommand \EOS [0]{\spacefactor3000\relax}%
\providecommand \BibitemShut  [1]{\csname bibitem#1\endcsname}%
\let\auto@bib@innerbib\@empty
\bibitem [{\citenamefont {Wen}(2007)}]{Wen2007textbook}%
  \BibitemOpen
  \bibfield  {author} {\bibinfo {author} {\bibfnamefont {X.-G.}\ \bibnamefont {Wen}},\ }\href {https://doi.org/10.1093/acprof:oso/9780199227259.001.0001} {\emph {\bibinfo {title} {Quantum Field Theory of Many-Body Systems: From the Origin of Sound to an Origin of Light and Electrons}}}\ (\bibinfo  {publisher} {Oxford University Press},\ \bibinfo {year} {2007})\BibitemShut {NoStop}%
\bibitem [{\citenamefont {Fradkin}(2013)}]{Fradkin2013textbook}%
  \BibitemOpen
  \bibfield  {author} {\bibinfo {author} {\bibfnamefont {E.}~\bibnamefont {Fradkin}},\ }\href@noop {} {\emph {\bibinfo {title} {Field Theories of Condensed Matter Physics}}},\ \bibinfo {edition} {2nd}\ ed.\ (\bibinfo  {publisher} {Cambridge University Press},\ \bibinfo {year} {2013})\BibitemShut {NoStop}%
\bibitem [{\citenamefont {Laughlin}(1981)}]{Laughlin1981prb}%
  \BibitemOpen
  \bibfield  {author} {\bibinfo {author} {\bibfnamefont {R.~B.}\ \bibnamefont {Laughlin}},\ }\bibfield  {title} {\bibinfo {title} {Quantized hall conductivity in two dimensions},\ }\href {https://doi.org/10.1103/PhysRevB.23.5632} {\bibfield  {journal} {\bibinfo  {journal} {Phys. Rev. B}\ }\textbf {\bibinfo {volume} {23}},\ \bibinfo {pages} {5632} (\bibinfo {year} {1981})}\BibitemShut {NoStop}%
\bibitem [{\citenamefont {Thouless}\ \emph {et~al.}(1982)\citenamefont {Thouless}, \citenamefont {Kohmoto}, \citenamefont {Nightingale},\ and\ \citenamefont {den Nijs}}]{Thouless1982prl}%
  \BibitemOpen
  \bibfield  {author} {\bibinfo {author} {\bibfnamefont {D.~J.}\ \bibnamefont {Thouless}}, \bibinfo {author} {\bibfnamefont {M.}~\bibnamefont {Kohmoto}}, \bibinfo {author} {\bibfnamefont {M.~P.}\ \bibnamefont {Nightingale}},\ and\ \bibinfo {author} {\bibfnamefont {M.}~\bibnamefont {den Nijs}},\ }\bibfield  {title} {\bibinfo {title} {Quantized hall conductance in a two-dimensional periodic potential},\ }\href {https://doi.org/10.1103/PhysRevLett.49.405} {\bibfield  {journal} {\bibinfo  {journal} {Phys. Rev. Lett.}\ }\textbf {\bibinfo {volume} {49}},\ \bibinfo {pages} {405} (\bibinfo {year} {1982})}\BibitemShut {NoStop}%
\bibitem [{\citenamefont {Savary}\ and\ \citenamefont {Balents}(2016)}]{Savary2017rpp}%
  \BibitemOpen
  \bibfield  {author} {\bibinfo {author} {\bibfnamefont {L.}~\bibnamefont {Savary}}\ and\ \bibinfo {author} {\bibfnamefont {L.}~\bibnamefont {Balents}},\ }\bibfield  {title} {\bibinfo {title} {Quantum spin liquids: a review},\ }\href {https://doi.org/10.1088/0034-4885/80/1/016502} {\bibfield  {journal} {\bibinfo  {journal} {Reports on Progress in Physics}\ }\textbf {\bibinfo {volume} {80}},\ \bibinfo {pages} {016502} (\bibinfo {year} {2016})}\BibitemShut {NoStop}%
\bibitem [{\citenamefont {Zhou}\ \emph {et~al.}(2017)\citenamefont {Zhou}, \citenamefont {Kanoda},\ and\ \citenamefont {Ng}}]{Zhou2017rmp}%
  \BibitemOpen
  \bibfield  {author} {\bibinfo {author} {\bibfnamefont {Y.}~\bibnamefont {Zhou}}, \bibinfo {author} {\bibfnamefont {K.}~\bibnamefont {Kanoda}},\ and\ \bibinfo {author} {\bibfnamefont {T.-K.}\ \bibnamefont {Ng}},\ }\bibfield  {title} {\bibinfo {title} {Quantum spin liquid states},\ }\href {https://doi.org/10.1103/RevModPhys.89.025003} {\bibfield  {journal} {\bibinfo  {journal} {Rev. Mod. Phys.}\ }\textbf {\bibinfo {volume} {89}},\ \bibinfo {pages} {025003} (\bibinfo {year} {2017})}\BibitemShut {NoStop}%
\bibitem [{\citenamefont {Broholm}\ \emph {et~al.}(2020)\citenamefont {Broholm}, \citenamefont {Cava}, \citenamefont {Kivelson}, \citenamefont {Nocera}, \citenamefont {Norman},\ and\ \citenamefont {Senthil}}]{Broholm2020science}%
  \BibitemOpen
  \bibfield  {author} {\bibinfo {author} {\bibfnamefont {C.}~\bibnamefont {Broholm}}, \bibinfo {author} {\bibfnamefont {R.~J.}\ \bibnamefont {Cava}}, \bibinfo {author} {\bibfnamefont {S.~A.}\ \bibnamefont {Kivelson}}, \bibinfo {author} {\bibfnamefont {D.~G.}\ \bibnamefont {Nocera}}, \bibinfo {author} {\bibfnamefont {M.~R.}\ \bibnamefont {Norman}},\ and\ \bibinfo {author} {\bibfnamefont {T.}~\bibnamefont {Senthil}},\ }\bibfield  {title} {\bibinfo {title} {Quantum spin liquids},\ }\href {https://doi.org/10.1126/science.aay0668} {\bibfield  {journal} {\bibinfo  {journal} {Science}\ }\textbf {\bibinfo {volume} {367}},\ \bibinfo {pages} {eaay0668} (\bibinfo {year} {2020})}\BibitemShut {NoStop}%
\bibitem [{\citenamefont {Qi}\ and\ \citenamefont {Zhang}(2011)}]{Qi2011rmp}%
  \BibitemOpen
  \bibfield  {author} {\bibinfo {author} {\bibfnamefont {X.-L.}\ \bibnamefont {Qi}}\ and\ \bibinfo {author} {\bibfnamefont {S.-C.}\ \bibnamefont {Zhang}},\ }\bibfield  {title} {\bibinfo {title} {Topological insulators and superconductors},\ }\href {https://doi.org/10.1103/RevModPhys.83.1057} {\bibfield  {journal} {\bibinfo  {journal} {Rev. Mod. Phys.}\ }\textbf {\bibinfo {volume} {83}},\ \bibinfo {pages} {1057} (\bibinfo {year} {2011})}\BibitemShut {NoStop}%
\bibitem [{\citenamefont {Tanaka}\ \emph {et~al.}(2012)\citenamefont {Tanaka}, \citenamefont {Sato},\ and\ \citenamefont {Nagaosa}}]{Tanaka2012jpsj}%
  \BibitemOpen
  \bibfield  {author} {\bibinfo {author} {\bibfnamefont {Y.}~\bibnamefont {Tanaka}}, \bibinfo {author} {\bibfnamefont {M.}~\bibnamefont {Sato}},\ and\ \bibinfo {author} {\bibfnamefont {N.}~\bibnamefont {Nagaosa}},\ }\bibfield  {title} {\bibinfo {title} {Symmetry and topology in superconductors –odd-frequency pairing and edge states–},\ }\href {https://doi.org/10.1143/JPSJ.81.011013} {\bibfield  {journal} {\bibinfo  {journal} {Journal of the Physical Society of Japan}\ }\textbf {\bibinfo {volume} {81}},\ \bibinfo {pages} {011013} (\bibinfo {year} {2012})}\BibitemShut {NoStop}%
\bibitem [{\citenamefont {Sato}\ and\ \citenamefont {Fujimoto}(2016)}]{Sato2016jpsj}%
  \BibitemOpen
  \bibfield  {author} {\bibinfo {author} {\bibfnamefont {M.}~\bibnamefont {Sato}}\ and\ \bibinfo {author} {\bibfnamefont {S.}~\bibnamefont {Fujimoto}},\ }\bibfield  {title} {\bibinfo {title} {Majorana fermions and topology in superconductors},\ }\href {https://doi.org/10.7566/JPSJ.85.072001} {\bibfield  {journal} {\bibinfo  {journal} {Journal of the Physical Society of Japan}\ }\textbf {\bibinfo {volume} {85}},\ \bibinfo {pages} {072001} (\bibinfo {year} {2016})}\BibitemShut {NoStop}%
\bibitem [{\citenamefont {Sato}\ and\ \citenamefont {Ando}(2017)}]{Sato2017rpp}%
  \BibitemOpen
  \bibfield  {author} {\bibinfo {author} {\bibfnamefont {M.}~\bibnamefont {Sato}}\ and\ \bibinfo {author} {\bibfnamefont {Y.}~\bibnamefont {Ando}},\ }\bibfield  {title} {\bibinfo {title} {Topological superconductors: a review},\ }\href {https://doi.org/10.1088/1361-6633/aa6ac7} {\bibfield  {journal} {\bibinfo  {journal} {Reports on Progress in Physics}\ }\textbf {\bibinfo {volume} {80}},\ \bibinfo {pages} {076501} (\bibinfo {year} {2017})}\BibitemShut {NoStop}%
\bibitem [{\citenamefont {Kitaev}(2001)}]{Kitaev2001pu}%
  \BibitemOpen
  \bibfield  {author} {\bibinfo {author} {\bibfnamefont {A.~Y.}\ \bibnamefont {Kitaev}},\ }\bibfield  {title} {\bibinfo {title} {Unpaired majorana fermions in quantum wires},\ }\href {https://doi.org/10.1070/1063-7869/44/10S/S29} {\bibfield  {journal} {\bibinfo  {journal} {Physics-Uspekhi}\ }\textbf {\bibinfo {volume} {44}},\ \bibinfo {pages} {131} (\bibinfo {year} {2001})}\BibitemShut {NoStop}%
\bibitem [{\citenamefont {Nayak}\ \emph {et~al.}(2008)\citenamefont {Nayak}, \citenamefont {Simon}, \citenamefont {Stern}, \citenamefont {Freedman},\ and\ \citenamefont {Das~Sarma}}]{Nayak2008rmp}%
  \BibitemOpen
  \bibfield  {author} {\bibinfo {author} {\bibfnamefont {C.}~\bibnamefont {Nayak}}, \bibinfo {author} {\bibfnamefont {S.~H.}\ \bibnamefont {Simon}}, \bibinfo {author} {\bibfnamefont {A.}~\bibnamefont {Stern}}, \bibinfo {author} {\bibfnamefont {M.}~\bibnamefont {Freedman}},\ and\ \bibinfo {author} {\bibfnamefont {S.}~\bibnamefont {Das~Sarma}},\ }\bibfield  {title} {\bibinfo {title} {Non-abelian anyons and topological quantum computation},\ }\href {https://doi.org/10.1103/RevModPhys.80.1083} {\bibfield  {journal} {\bibinfo  {journal} {Rev. Mod. Phys.}\ }\textbf {\bibinfo {volume} {80}},\ \bibinfo {pages} {1083} (\bibinfo {year} {2008})}\BibitemShut {NoStop}%
\bibitem [{\citenamefont {Moriya}\ and\ \citenamefont {Ueda}(2000)}]{Moriya2000ap}%
  \BibitemOpen
  \bibfield  {author} {\bibinfo {author} {\bibfnamefont {T.}~\bibnamefont {Moriya}}\ and\ \bibinfo {author} {\bibfnamefont {K.}~\bibnamefont {Ueda}},\ }\bibfield  {title} {\bibinfo {title} {Spin fluctuations and high temperature superconductivity},\ }\href {https://doi.org/10.1080/000187300412248} {\bibfield  {journal} {\bibinfo  {journal} {Advances in Physics}\ }\textbf {\bibinfo {volume} {49}},\ \bibinfo {pages} {555} (\bibinfo {year} {2000})}\BibitemShut {NoStop}%
\bibitem [{\citenamefont {Yanase}\ \emph {et~al.}(2003)\citenamefont {Yanase}, \citenamefont {Jujo}, \citenamefont {Nomura}, \citenamefont {Ikeda}, \citenamefont {Hotta},\ and\ \citenamefont {Yamada}}]{Yanase2003pr}%
  \BibitemOpen
  \bibfield  {author} {\bibinfo {author} {\bibfnamefont {Y.}~\bibnamefont {Yanase}}, \bibinfo {author} {\bibfnamefont {T.}~\bibnamefont {Jujo}}, \bibinfo {author} {\bibfnamefont {T.}~\bibnamefont {Nomura}}, \bibinfo {author} {\bibfnamefont {H.}~\bibnamefont {Ikeda}}, \bibinfo {author} {\bibfnamefont {T.}~\bibnamefont {Hotta}},\ and\ \bibinfo {author} {\bibfnamefont {K.}~\bibnamefont {Yamada}},\ }\bibfield  {title} {\bibinfo {title} {Theory of superconductivity in strongly correlated electron systems},\ }\href {https://doi.org/https://doi.org/10.1016/j.physrep.2003.07.002} {\bibfield  {journal} {\bibinfo  {journal} {Physics Reports}\ }\textbf {\bibinfo {volume} {387}},\ \bibinfo {pages} {1} (\bibinfo {year} {2003})}\BibitemShut {NoStop}%
\bibitem [{\citenamefont {Aoki}\ \emph {et~al.}(2022)\citenamefont {Aoki}, \citenamefont {Brison}, \citenamefont {Flouquet}, \citenamefont {Ishida}, \citenamefont {Knebel}, \citenamefont {Tokunaga},\ and\ \citenamefont {Yanase}}]{Aoki2022jpcm}%
  \BibitemOpen
  \bibfield  {author} {\bibinfo {author} {\bibfnamefont {D.}~\bibnamefont {Aoki}}, \bibinfo {author} {\bibfnamefont {J.-P.}\ \bibnamefont {Brison}}, \bibinfo {author} {\bibfnamefont {J.}~\bibnamefont {Flouquet}}, \bibinfo {author} {\bibfnamefont {K.}~\bibnamefont {Ishida}}, \bibinfo {author} {\bibfnamefont {G.}~\bibnamefont {Knebel}}, \bibinfo {author} {\bibfnamefont {Y.}~\bibnamefont {Tokunaga}},\ and\ \bibinfo {author} {\bibfnamefont {Y.}~\bibnamefont {Yanase}},\ }\bibfield  {title} {\bibinfo {title} {Unconventional superconductivity in ute2},\ }\href {https://doi.org/10.1088/1361-648X/ac5863} {\bibfield  {journal} {\bibinfo  {journal} {Journal of Physics: Condensed Matter}\ }\textbf {\bibinfo {volume} {34}},\ \bibinfo {pages} {243002} (\bibinfo {year} {2022})}\BibitemShut {NoStop}%
\bibitem [{\citenamefont {Fischer}\ \emph {et~al.}(2011)\citenamefont {Fischer}, \citenamefont {Loder},\ and\ \citenamefont {Sigrist}}]{Fischer2011prb}%
  \BibitemOpen
  \bibfield  {author} {\bibinfo {author} {\bibfnamefont {M.~H.}\ \bibnamefont {Fischer}}, \bibinfo {author} {\bibfnamefont {F.}~\bibnamefont {Loder}},\ and\ \bibinfo {author} {\bibfnamefont {M.}~\bibnamefont {Sigrist}},\ }\bibfield  {title} {\bibinfo {title} {Superconductivity and local noncentrosymmetricity in crystal lattices},\ }\href {https://doi.org/10.1103/PhysRevB.84.184533} {\bibfield  {journal} {\bibinfo  {journal} {Phys. Rev. B}\ }\textbf {\bibinfo {volume} {84}},\ \bibinfo {pages} {184533} (\bibinfo {year} {2011})}\BibitemShut {NoStop}%
\bibitem [{\citenamefont {Maruyama}\ \emph {et~al.}(2012)\citenamefont {Maruyama}, \citenamefont {Sigrist},\ and\ \citenamefont {Yanase}}]{Maruyama2012jpsj}%
  \BibitemOpen
  \bibfield  {author} {\bibinfo {author} {\bibfnamefont {D.}~\bibnamefont {Maruyama}}, \bibinfo {author} {\bibfnamefont {M.}~\bibnamefont {Sigrist}},\ and\ \bibinfo {author} {\bibfnamefont {Y.}~\bibnamefont {Yanase}},\ }\bibfield  {title} {\bibinfo {title} {Locally non-centrosymmetric superconductivity in multilayer systems},\ }\href {https://doi.org/10.1143/JPSJ.81.034702} {\bibfield  {journal} {\bibinfo  {journal} {Journal of the Physical Society of Japan}\ }\textbf {\bibinfo {volume} {81}},\ \bibinfo {pages} {034702} (\bibinfo {year} {2012})}\BibitemShut {NoStop}%
\bibitem [{\citenamefont {Yoshida}\ \emph {et~al.}(2012)\citenamefont {Yoshida}, \citenamefont {Sigrist},\ and\ \citenamefont {Yanase}}]{Yoshida2012prb}%
  \BibitemOpen
  \bibfield  {author} {\bibinfo {author} {\bibfnamefont {T.}~\bibnamefont {Yoshida}}, \bibinfo {author} {\bibfnamefont {M.}~\bibnamefont {Sigrist}},\ and\ \bibinfo {author} {\bibfnamefont {Y.}~\bibnamefont {Yanase}},\ }\bibfield  {title} {\bibinfo {title} {Pair-density wave states through spin-orbit coupling in multilayer superconductors},\ }\href {https://doi.org/10.1103/PhysRevB.86.134514} {\bibfield  {journal} {\bibinfo  {journal} {Phys. Rev. B}\ }\textbf {\bibinfo {volume} {86}},\ \bibinfo {pages} {134514} (\bibinfo {year} {2012})}\BibitemShut {NoStop}%
\bibitem [{\citenamefont {Fischer}\ \emph {et~al.}(2023)\citenamefont {Fischer}, \citenamefont {Sigrist}, \citenamefont {Agterberg},\ and\ \citenamefont {Yanase}}]{Fischer2023arcmp}%
  \BibitemOpen
  \bibfield  {author} {\bibinfo {author} {\bibfnamefont {M.~H.}\ \bibnamefont {Fischer}}, \bibinfo {author} {\bibfnamefont {M.}~\bibnamefont {Sigrist}}, \bibinfo {author} {\bibfnamefont {D.~F.}\ \bibnamefont {Agterberg}},\ and\ \bibinfo {author} {\bibfnamefont {Y.}~\bibnamefont {Yanase}},\ }\bibfield  {title} {\bibinfo {title} {Superconductivity and local inversion-symmetry breaking},\ }\href {https://doi.org/https://doi.org/10.1146/annurev-conmatphys-040521-042511} {\bibfield  {journal} {\bibinfo  {journal} {Annual Review of Condensed Matter Physics}\ }\textbf {\bibinfo {volume} {14}},\ \bibinfo {pages} {153} (\bibinfo {year} {2023})}\BibitemShut {NoStop}%
\bibitem [{\citenamefont {M\"ockli}\ \emph {et~al.}(2018)\citenamefont {M\"ockli}, \citenamefont {Yanase},\ and\ \citenamefont {Sigrist}}]{Mockli2018prb}%
  \BibitemOpen
  \bibfield  {author} {\bibinfo {author} {\bibfnamefont {D.}~\bibnamefont {M\"ockli}}, \bibinfo {author} {\bibfnamefont {Y.}~\bibnamefont {Yanase}},\ and\ \bibinfo {author} {\bibfnamefont {M.}~\bibnamefont {Sigrist}},\ }\bibfield  {title} {\bibinfo {title} {Orbitally limited pair-density-wave phase of multilayer superconductors},\ }\href {https://doi.org/10.1103/PhysRevB.97.144508} {\bibfield  {journal} {\bibinfo  {journal} {Phys. Rev. B}\ }\textbf {\bibinfo {volume} {97}},\ \bibinfo {pages} {144508} (\bibinfo {year} {2018})}\BibitemShut {NoStop}%
\bibitem [{\citenamefont {M\"ockli}\ and\ \citenamefont {Ramires}(2021)}]{Mockli2021prb}%
  \BibitemOpen
  \bibfield  {author} {\bibinfo {author} {\bibfnamefont {D.}~\bibnamefont {M\"ockli}}\ and\ \bibinfo {author} {\bibfnamefont {A.}~\bibnamefont {Ramires}},\ }\bibfield  {title} {\bibinfo {title} {Superconductivity in disordered locally noncentrosymmetric materials: An application to ${\mathrm{cerh}}_{2}{\mathrm{as}}_{2}$},\ }\href {https://doi.org/10.1103/PhysRevB.104.134517} {\bibfield  {journal} {\bibinfo  {journal} {Phys. Rev. B}\ }\textbf {\bibinfo {volume} {104}},\ \bibinfo {pages} {134517} (\bibinfo {year} {2021})}\BibitemShut {NoStop}%
\bibitem [{\citenamefont {Maki}(1964)}]{Maki1964ppf}%
  \BibitemOpen
  \bibfield  {author} {\bibinfo {author} {\bibfnamefont {K.}~\bibnamefont {Maki}},\ }\bibfield  {title} {\bibinfo {title} {The magnetic properties of superconducting alloys. ii},\ }\href {https://doi.org/10.1103/PhysicsPhysiqueFizika.1.127} {\bibfield  {journal} {\bibinfo  {journal} {Physics Physique Fizika}\ }\textbf {\bibinfo {volume} {1}},\ \bibinfo {pages} {127} (\bibinfo {year} {1964})}\BibitemShut {NoStop}%
\bibitem [{\citenamefont {Khim}\ \emph {et~al.}(2021)\citenamefont {Khim}, \citenamefont {Landaeta}, \citenamefont {Banda}, \citenamefont {Bannor}, \citenamefont {Brando}, \citenamefont {Brydon}, \citenamefont {Hafner}, \citenamefont {Küchler}, \citenamefont {Cardoso-Gil}, \citenamefont {Stockert}, \citenamefont {Mackenzie}, \citenamefont {Agterberg}, \citenamefont {Geibel},\ and\ \citenamefont {Hassinger}}]{Khim2021science}%
  \BibitemOpen
  \bibfield  {author} {\bibinfo {author} {\bibfnamefont {S.}~\bibnamefont {Khim}}, \bibinfo {author} {\bibfnamefont {J.~F.}\ \bibnamefont {Landaeta}}, \bibinfo {author} {\bibfnamefont {J.}~\bibnamefont {Banda}}, \bibinfo {author} {\bibfnamefont {N.}~\bibnamefont {Bannor}}, \bibinfo {author} {\bibfnamefont {M.}~\bibnamefont {Brando}}, \bibinfo {author} {\bibfnamefont {P.~M.~R.}\ \bibnamefont {Brydon}}, \bibinfo {author} {\bibfnamefont {D.}~\bibnamefont {Hafner}}, \bibinfo {author} {\bibfnamefont {R.}~\bibnamefont {Küchler}}, \bibinfo {author} {\bibfnamefont {R.}~\bibnamefont {Cardoso-Gil}}, \bibinfo {author} {\bibfnamefont {U.}~\bibnamefont {Stockert}}, \bibinfo {author} {\bibfnamefont {A.~P.}\ \bibnamefont {Mackenzie}}, \bibinfo {author} {\bibfnamefont {D.~F.}\ \bibnamefont {Agterberg}}, \bibinfo {author} {\bibfnamefont {C.}~\bibnamefont {Geibel}},\ and\ \bibinfo {author} {\bibfnamefont {E.}~\bibnamefont {Hassinger}},\ }\bibfield  {title} {\bibinfo {title} {Field-induced transition within the superconducting
  state of cerh<sub>2</sub>as<sub>2</sub>},\ }\href {https://doi.org/10.1126/science.abe7518} {\bibfield  {journal} {\bibinfo  {journal} {Science}\ }\textbf {\bibinfo {volume} {373}},\ \bibinfo {pages} {1012} (\bibinfo {year} {2021})}\BibitemShut {NoStop}%
\bibitem [{\citenamefont {Kimura}\ \emph {et~al.}(2021)\citenamefont {Kimura}, \citenamefont {Sichelschmidt},\ and\ \citenamefont {Khim}}]{Kimura2021prb}%
  \BibitemOpen
  \bibfield  {author} {\bibinfo {author} {\bibfnamefont {S.-i.}\ \bibnamefont {Kimura}}, \bibinfo {author} {\bibfnamefont {J.}~\bibnamefont {Sichelschmidt}},\ and\ \bibinfo {author} {\bibfnamefont {S.}~\bibnamefont {Khim}},\ }\bibfield  {title} {\bibinfo {title} {Optical study of the electronic structure of locally noncentrosymmetric $\mathrm{Ce}{\mathrm{rh}}_{2}{\mathrm{as}}_{2}$},\ }\href {https://doi.org/10.1103/PhysRevB.104.245116} {\bibfield  {journal} {\bibinfo  {journal} {Phys. Rev. B}\ }\textbf {\bibinfo {volume} {104}},\ \bibinfo {pages} {245116} (\bibinfo {year} {2021})}\BibitemShut {NoStop}%
\bibitem [{\citenamefont {Onishi}\ \emph {et~al.}(2022)\citenamefont {Onishi}, \citenamefont {Stockert}, \citenamefont {Khim}, \citenamefont {Banda}, \citenamefont {Brando},\ and\ \citenamefont {Hassinger}}]{Onisi2022fem}%
  \BibitemOpen
  \bibfield  {author} {\bibinfo {author} {\bibfnamefont {S.}~\bibnamefont {Onishi}}, \bibinfo {author} {\bibfnamefont {U.}~\bibnamefont {Stockert}}, \bibinfo {author} {\bibfnamefont {S.}~\bibnamefont {Khim}}, \bibinfo {author} {\bibfnamefont {J.}~\bibnamefont {Banda}}, \bibinfo {author} {\bibfnamefont {M.}~\bibnamefont {Brando}},\ and\ \bibinfo {author} {\bibfnamefont {E.}~\bibnamefont {Hassinger}},\ }\bibfield  {title} {\bibinfo {title} {Low-temperature thermal conductivity of the two-phase superconductor cerh2as2},\ }\bibfield  {journal} {\bibinfo  {journal} {Frontiers in Electronic Materials}\ }\textbf {\bibinfo {volume} {Volume 2 - 2022}},\ \href {https://doi.org/10.3389/femat.2022.880579} {10.3389/femat.2022.880579} (\bibinfo {year} {2022})\BibitemShut {NoStop}%
\bibitem [{\citenamefont {Kibune}\ \emph {et~al.}(2022)\citenamefont {Kibune}, \citenamefont {Kitagawa}, \citenamefont {Kinjo}, \citenamefont {Ogata}, \citenamefont {Manago}, \citenamefont {Taniguchi}, \citenamefont {Ishida}, \citenamefont {Brando}, \citenamefont {Hassinger}, \citenamefont {Rosner}, \citenamefont {Geibel},\ and\ \citenamefont {Khim}}]{Kibune2022prl}%
  \BibitemOpen
  \bibfield  {author} {\bibinfo {author} {\bibfnamefont {M.}~\bibnamefont {Kibune}}, \bibinfo {author} {\bibfnamefont {S.}~\bibnamefont {Kitagawa}}, \bibinfo {author} {\bibfnamefont {K.}~\bibnamefont {Kinjo}}, \bibinfo {author} {\bibfnamefont {S.}~\bibnamefont {Ogata}}, \bibinfo {author} {\bibfnamefont {M.}~\bibnamefont {Manago}}, \bibinfo {author} {\bibfnamefont {T.}~\bibnamefont {Taniguchi}}, \bibinfo {author} {\bibfnamefont {K.}~\bibnamefont {Ishida}}, \bibinfo {author} {\bibfnamefont {M.}~\bibnamefont {Brando}}, \bibinfo {author} {\bibfnamefont {E.}~\bibnamefont {Hassinger}}, \bibinfo {author} {\bibfnamefont {H.}~\bibnamefont {Rosner}}, \bibinfo {author} {\bibfnamefont {C.}~\bibnamefont {Geibel}},\ and\ \bibinfo {author} {\bibfnamefont {S.}~\bibnamefont {Khim}},\ }\bibfield  {title} {\bibinfo {title} {Observation of antiferromagnetic order as odd-parity multipoles inside the superconducting phase in {${\mathrm{CeRh}}_{2}{\mathrm{As}}_{2}$}},\ }\href {https://doi.org/10.1103/PhysRevLett.128.057002}
  {\bibfield  {journal} {\bibinfo  {journal} {Phys. Rev. Lett.}\ }\textbf {\bibinfo {volume} {128}},\ \bibinfo {pages} {057002} (\bibinfo {year} {2022})}\BibitemShut {NoStop}%
\bibitem [{\citenamefont {Hafner}\ \emph {et~al.}(2022)\citenamefont {Hafner}, \citenamefont {Khanenko}, \citenamefont {Eljaouhari}, \citenamefont {K\"uchler}, \citenamefont {Banda}, \citenamefont {Bannor}, \citenamefont {L\"uhmann}, \citenamefont {Landaeta}, \citenamefont {Mishra}, \citenamefont {Sheikin}, \citenamefont {Hassinger}, \citenamefont {Khim}, \citenamefont {Geibel}, \citenamefont {Zwicknagl},\ and\ \citenamefont {Brando}}]{Hafner2022prx}%
  \BibitemOpen
  \bibfield  {author} {\bibinfo {author} {\bibfnamefont {D.}~\bibnamefont {Hafner}}, \bibinfo {author} {\bibfnamefont {P.}~\bibnamefont {Khanenko}}, \bibinfo {author} {\bibfnamefont {E.-O.}\ \bibnamefont {Eljaouhari}}, \bibinfo {author} {\bibfnamefont {R.}~\bibnamefont {K\"uchler}}, \bibinfo {author} {\bibfnamefont {J.}~\bibnamefont {Banda}}, \bibinfo {author} {\bibfnamefont {N.}~\bibnamefont {Bannor}}, \bibinfo {author} {\bibfnamefont {T.}~\bibnamefont {L\"uhmann}}, \bibinfo {author} {\bibfnamefont {J.~F.}\ \bibnamefont {Landaeta}}, \bibinfo {author} {\bibfnamefont {S.}~\bibnamefont {Mishra}}, \bibinfo {author} {\bibfnamefont {I.}~\bibnamefont {Sheikin}}, \bibinfo {author} {\bibfnamefont {E.}~\bibnamefont {Hassinger}}, \bibinfo {author} {\bibfnamefont {S.}~\bibnamefont {Khim}}, \bibinfo {author} {\bibfnamefont {C.}~\bibnamefont {Geibel}}, \bibinfo {author} {\bibfnamefont {G.}~\bibnamefont {Zwicknagl}},\ and\ \bibinfo {author} {\bibfnamefont {M.}~\bibnamefont {Brando}},\ }\bibfield  {title} {\bibinfo {title}
  {Possible quadrupole density wave in the superconducting kondo lattice ${\mathrm{cerh}}_{2}{\mathrm{as}}_{2}$},\ }\href {https://doi.org/10.1103/PhysRevX.12.011023} {\bibfield  {journal} {\bibinfo  {journal} {Phys. Rev. X}\ }\textbf {\bibinfo {volume} {12}},\ \bibinfo {pages} {011023} (\bibinfo {year} {2022})}\BibitemShut {NoStop}%
\bibitem [{\citenamefont {Kitagawa}\ \emph {et~al.}(2022)\citenamefont {Kitagawa}, \citenamefont {Kibune}, \citenamefont {Kinjo}, \citenamefont {Manago}, \citenamefont {Taniguchi}, \citenamefont {Ishida}, \citenamefont {Brando}, \citenamefont {Hassinger}, \citenamefont {Geibel},\ and\ \citenamefont {Khim}}]{Kitagawa2022jpsj}%
  \BibitemOpen
  \bibfield  {author} {\bibinfo {author} {\bibfnamefont {S.}~\bibnamefont {Kitagawa}}, \bibinfo {author} {\bibfnamefont {M.}~\bibnamefont {Kibune}}, \bibinfo {author} {\bibfnamefont {K.}~\bibnamefont {Kinjo}}, \bibinfo {author} {\bibfnamefont {M.}~\bibnamefont {Manago}}, \bibinfo {author} {\bibfnamefont {T.}~\bibnamefont {Taniguchi}}, \bibinfo {author} {\bibfnamefont {K.}~\bibnamefont {Ishida}}, \bibinfo {author} {\bibfnamefont {M.}~\bibnamefont {Brando}}, \bibinfo {author} {\bibfnamefont {E.}~\bibnamefont {Hassinger}}, \bibinfo {author} {\bibfnamefont {C.}~\bibnamefont {Geibel}},\ and\ \bibinfo {author} {\bibfnamefont {S.}~\bibnamefont {Khim}},\ }\bibfield  {title} {\bibinfo {title} {Two-dimensional {XY}-type magnetic properties of locally noncentrosymmetric superconductor {CeRh$_2$As$_2$}},\ }\href {https://doi.org/10.7566/JPSJ.91.043702} {\bibfield  {journal} {\bibinfo  {journal} {Journal of the Physical Society of Japan}\ }\textbf {\bibinfo {volume} {91}},\ \bibinfo {pages} {043702} (\bibinfo {year}
  {2022})}\BibitemShut {NoStop}%
\bibitem [{\citenamefont {Landaeta}\ \emph {et~al.}(2022)\citenamefont {Landaeta}, \citenamefont {Khanenko}, \citenamefont {Cavanagh}, \citenamefont {Geibel}, \citenamefont {Khim}, \citenamefont {Mishra}, \citenamefont {Sheikin}, \citenamefont {Brydon}, \citenamefont {Agterberg}, \citenamefont {Brando},\ and\ \citenamefont {Hassinger}}]{Landaeta2022prx}%
  \BibitemOpen
  \bibfield  {author} {\bibinfo {author} {\bibfnamefont {J.~F.}\ \bibnamefont {Landaeta}}, \bibinfo {author} {\bibfnamefont {P.}~\bibnamefont {Khanenko}}, \bibinfo {author} {\bibfnamefont {D.~C.}\ \bibnamefont {Cavanagh}}, \bibinfo {author} {\bibfnamefont {C.}~\bibnamefont {Geibel}}, \bibinfo {author} {\bibfnamefont {S.}~\bibnamefont {Khim}}, \bibinfo {author} {\bibfnamefont {S.}~\bibnamefont {Mishra}}, \bibinfo {author} {\bibfnamefont {I.}~\bibnamefont {Sheikin}}, \bibinfo {author} {\bibfnamefont {P.~M.~R.}\ \bibnamefont {Brydon}}, \bibinfo {author} {\bibfnamefont {D.~F.}\ \bibnamefont {Agterberg}}, \bibinfo {author} {\bibfnamefont {M.}~\bibnamefont {Brando}},\ and\ \bibinfo {author} {\bibfnamefont {E.}~\bibnamefont {Hassinger}},\ }\bibfield  {title} {\bibinfo {title} {Field-angle dependence reveals odd-parity superconductivity in ${\mathrm{cerh}}_{2}{\mathrm{as}}_{2}$},\ }\href {https://doi.org/10.1103/PhysRevX.12.031001} {\bibfield  {journal} {\bibinfo  {journal} {Phys. Rev. X}\ }\textbf {\bibinfo {volume}
  {12}},\ \bibinfo {pages} {031001} (\bibinfo {year} {2022})}\BibitemShut {NoStop}%
\bibitem [{\citenamefont {Mishra}\ \emph {et~al.}(2022)\citenamefont {Mishra}, \citenamefont {Liu}, \citenamefont {Bauer}, \citenamefont {Ronning},\ and\ \citenamefont {Thomas}}]{Mishra2022prb}%
  \BibitemOpen
  \bibfield  {author} {\bibinfo {author} {\bibfnamefont {S.}~\bibnamefont {Mishra}}, \bibinfo {author} {\bibfnamefont {Y.}~\bibnamefont {Liu}}, \bibinfo {author} {\bibfnamefont {E.~D.}\ \bibnamefont {Bauer}}, \bibinfo {author} {\bibfnamefont {F.}~\bibnamefont {Ronning}},\ and\ \bibinfo {author} {\bibfnamefont {S.~M.}\ \bibnamefont {Thomas}},\ }\bibfield  {title} {\bibinfo {title} {Anisotropic magnetotransport properties of the heavy-fermion superconductor ${\mathrm{cerh}}_{2}{\mathrm{as}}_{2}$},\ }\href {https://doi.org/10.1103/PhysRevB.106.L140502} {\bibfield  {journal} {\bibinfo  {journal} {Phys. Rev. B}\ }\textbf {\bibinfo {volume} {106}},\ \bibinfo {pages} {L140502} (\bibinfo {year} {2022})}\BibitemShut {NoStop}%
\bibitem [{\citenamefont {Ogata}\ \emph {et~al.}(2023{\natexlab{a}})\citenamefont {Ogata}, \citenamefont {Kitagawa}, \citenamefont {Kinjo}, \citenamefont {Ishida}, \citenamefont {Brando}, \citenamefont {Hassinger}, \citenamefont {Geibel},\ and\ \citenamefont {Khim}}]{Ogata2023prl}%
  \BibitemOpen
  \bibfield  {author} {\bibinfo {author} {\bibfnamefont {S.}~\bibnamefont {Ogata}}, \bibinfo {author} {\bibfnamefont {S.}~\bibnamefont {Kitagawa}}, \bibinfo {author} {\bibfnamefont {K.}~\bibnamefont {Kinjo}}, \bibinfo {author} {\bibfnamefont {K.}~\bibnamefont {Ishida}}, \bibinfo {author} {\bibfnamefont {M.}~\bibnamefont {Brando}}, \bibinfo {author} {\bibfnamefont {E.}~\bibnamefont {Hassinger}}, \bibinfo {author} {\bibfnamefont {C.}~\bibnamefont {Geibel}},\ and\ \bibinfo {author} {\bibfnamefont {S.}~\bibnamefont {Khim}},\ }\bibfield  {title} {\bibinfo {title} {Parity transition of spin-singlet superconductivity using sublattice degrees of freedom},\ }\href {https://doi.org/10.1103/PhysRevLett.130.166001} {\bibfield  {journal} {\bibinfo  {journal} {Phys. Rev. Lett.}\ }\textbf {\bibinfo {volume} {130}},\ \bibinfo {pages} {166001} (\bibinfo {year} {2023}{\natexlab{a}})}\BibitemShut {NoStop}%
\bibitem [{\citenamefont {Semeniuk}\ \emph {et~al.}(2023)\citenamefont {Semeniuk}, \citenamefont {Hafner}, \citenamefont {Khanenko}, \citenamefont {L\"uhmann}, \citenamefont {Banda}, \citenamefont {Landaeta}, \citenamefont {Geibel}, \citenamefont {Khim}, \citenamefont {Hassinger},\ and\ \citenamefont {Brando}}]{Semeniuk2023prb}%
  \BibitemOpen
  \bibfield  {author} {\bibinfo {author} {\bibfnamefont {K.}~\bibnamefont {Semeniuk}}, \bibinfo {author} {\bibfnamefont {D.}~\bibnamefont {Hafner}}, \bibinfo {author} {\bibfnamefont {P.}~\bibnamefont {Khanenko}}, \bibinfo {author} {\bibfnamefont {T.}~\bibnamefont {L\"uhmann}}, \bibinfo {author} {\bibfnamefont {J.}~\bibnamefont {Banda}}, \bibinfo {author} {\bibfnamefont {J.~F.}\ \bibnamefont {Landaeta}}, \bibinfo {author} {\bibfnamefont {C.}~\bibnamefont {Geibel}}, \bibinfo {author} {\bibfnamefont {S.}~\bibnamefont {Khim}}, \bibinfo {author} {\bibfnamefont {E.}~\bibnamefont {Hassinger}},\ and\ \bibinfo {author} {\bibfnamefont {M.}~\bibnamefont {Brando}},\ }\bibfield  {title} {\bibinfo {title} {Decoupling multiphase superconductivity from normal state ordering in ${\mathrm{cerh}}_{2}{\mathrm{as}}_{2}$},\ }\href {https://doi.org/10.1103/PhysRevB.107.L220504} {\bibfield  {journal} {\bibinfo  {journal} {Phys. Rev. B}\ }\textbf {\bibinfo {volume} {107}},\ \bibinfo {pages} {L220504} (\bibinfo {year}
  {2023})}\BibitemShut {NoStop}%
\bibitem [{\citenamefont {Siddiquee}\ \emph {et~al.}(2023)\citenamefont {Siddiquee}, \citenamefont {Rehfuss}, \citenamefont {Broyles},\ and\ \citenamefont {Ran}}]{Siddiquee2023prb}%
  \BibitemOpen
  \bibfield  {author} {\bibinfo {author} {\bibfnamefont {H.}~\bibnamefont {Siddiquee}}, \bibinfo {author} {\bibfnamefont {Z.}~\bibnamefont {Rehfuss}}, \bibinfo {author} {\bibfnamefont {C.}~\bibnamefont {Broyles}},\ and\ \bibinfo {author} {\bibfnamefont {S.}~\bibnamefont {Ran}},\ }\bibfield  {title} {\bibinfo {title} {Pressure dependence of superconductivity in ${\mathrm{cerh}}_{2}{\mathrm{as}}_{2}$},\ }\href {https://doi.org/10.1103/PhysRevB.108.L020504} {\bibfield  {journal} {\bibinfo  {journal} {Phys. Rev. B}\ }\textbf {\bibinfo {volume} {108}},\ \bibinfo {pages} {L020504} (\bibinfo {year} {2023})}\BibitemShut {NoStop}%
\bibitem [{\citenamefont {Ogata}\ \emph {et~al.}(2023{\natexlab{b}})\citenamefont {Ogata}, \citenamefont {Kitagawa}, \citenamefont {Kibune}, \citenamefont {Ishida}, \citenamefont {Kinjo}, \citenamefont {Brando}, \citenamefont {Geibel}, \citenamefont {Khim},\ and\ \citenamefont {Hassinger}}]{Ogata2023npsm}%
  \BibitemOpen
  \bibfield  {author} {\bibinfo {author} {\bibfnamefont {S.}~\bibnamefont {Ogata}}, \bibinfo {author} {\bibfnamefont {S.}~\bibnamefont {Kitagawa}}, \bibinfo {author} {\bibfnamefont {M.}~\bibnamefont {Kibune}}, \bibinfo {author} {\bibfnamefont {K.}~\bibnamefont {Ishida}}, \bibinfo {author} {\bibfnamefont {K.}~\bibnamefont {Kinjo}}, \bibinfo {author} {\bibfnamefont {M.}~\bibnamefont {Brando}}, \bibinfo {author} {\bibfnamefont {C.}~\bibnamefont {Geibel}}, \bibinfo {author} {\bibfnamefont {S.}~\bibnamefont {Khim}},\ and\ \bibinfo {author} {\bibfnamefont {E.}~\bibnamefont {Hassinger}},\ }\bibfield  {title} {\bibinfo {title} {Investigation of the hyperfine coupling constant of locally noncentrosymmetric heavy-fermion superconductor {${\mathrm{CeRh}}_{2}{\mathrm{As}}_{2}$}},\ }\href {https://doi.org/10.3938/NPSM.73.1115} {\bibfield  {journal} {\bibinfo  {journal} {New Physics: Sae Mulli}\ }\textbf {\bibinfo {volume} {73}},\ \bibinfo {pages} {1115} (\bibinfo {year} {2023}{\natexlab{b}})}\BibitemShut {NoStop}%
\bibitem [{\citenamefont {Christovam}\ \emph {et~al.}(2024)\citenamefont {Christovam}, \citenamefont {Ferreira-Carvalho}, \citenamefont {Marino}, \citenamefont {Sundermann}, \citenamefont {Takegami}, \citenamefont {Melendez-Sans}, \citenamefont {Tsuei}, \citenamefont {Hu}, \citenamefont {R\"o\ss{}ler}, \citenamefont {Valvidares}, \citenamefont {Haverkort}, \citenamefont {Liu}, \citenamefont {Bauer}, \citenamefont {Tjeng}, \citenamefont {Zwicknagl},\ and\ \citenamefont {Severing}}]{Christovam2024prl}%
  \BibitemOpen
  \bibfield  {author} {\bibinfo {author} {\bibfnamefont {D.~S.}\ \bibnamefont {Christovam}}, \bibinfo {author} {\bibfnamefont {M.}~\bibnamefont {Ferreira-Carvalho}}, \bibinfo {author} {\bibfnamefont {A.}~\bibnamefont {Marino}}, \bibinfo {author} {\bibfnamefont {M.}~\bibnamefont {Sundermann}}, \bibinfo {author} {\bibfnamefont {D.}~\bibnamefont {Takegami}}, \bibinfo {author} {\bibfnamefont {A.}~\bibnamefont {Melendez-Sans}}, \bibinfo {author} {\bibfnamefont {K.~D.}\ \bibnamefont {Tsuei}}, \bibinfo {author} {\bibfnamefont {Z.}~\bibnamefont {Hu}}, \bibinfo {author} {\bibfnamefont {S.}~\bibnamefont {R\"o\ss{}ler}}, \bibinfo {author} {\bibfnamefont {M.}~\bibnamefont {Valvidares}}, \bibinfo {author} {\bibfnamefont {M.~W.}\ \bibnamefont {Haverkort}}, \bibinfo {author} {\bibfnamefont {Y.}~\bibnamefont {Liu}}, \bibinfo {author} {\bibfnamefont {E.~D.}\ \bibnamefont {Bauer}}, \bibinfo {author} {\bibfnamefont {L.~H.}\ \bibnamefont {Tjeng}}, \bibinfo {author} {\bibfnamefont {G.}~\bibnamefont {Zwicknagl}},\ and\ \bibinfo
  {author} {\bibfnamefont {A.}~\bibnamefont {Severing}},\ }\bibfield  {title} {\bibinfo {title} {Spectroscopic evidence of kondo-induced quasiquartet in ${\mathrm{cerh}}_{2}{\mathrm{as}}_{2}$},\ }\href {https://doi.org/10.1103/PhysRevLett.132.046401} {\bibfield  {journal} {\bibinfo  {journal} {Phys. Rev. Lett.}\ }\textbf {\bibinfo {volume} {132}},\ \bibinfo {pages} {046401} (\bibinfo {year} {2024})}\BibitemShut {NoStop}%
\bibitem [{\citenamefont {Chajewski}\ and\ \citenamefont {Kaczorowski}(2024)}]{Chajewski2024prl}%
  \BibitemOpen
  \bibfield  {author} {\bibinfo {author} {\bibfnamefont {G.}~\bibnamefont {Chajewski}}\ and\ \bibinfo {author} {\bibfnamefont {D.}~\bibnamefont {Kaczorowski}},\ }\bibfield  {title} {\bibinfo {title} {Discovery of magnetic phase transitions in heavy-fermion superconductor ${\mathrm{cerh}}_{2}{\mathrm{as}}_{2}$},\ }\href {https://doi.org/10.1103/PhysRevLett.132.076504} {\bibfield  {journal} {\bibinfo  {journal} {Phys. Rev. Lett.}\ }\textbf {\bibinfo {volume} {132}},\ \bibinfo {pages} {076504} (\bibinfo {year} {2024})}\BibitemShut {NoStop}%
\bibitem [{\citenamefont {Chen}\ \emph {et~al.}(2024{\natexlab{a}})\citenamefont {Chen}, \citenamefont {Wang}, \citenamefont {Ishizuka}, \citenamefont {Zhang}, \citenamefont {Nogaki}, \citenamefont {Cheng}, \citenamefont {Yang}, \citenamefont {Chen}, \citenamefont {Zhu}, \citenamefont {Liu}, \citenamefont {Mei}, \citenamefont {Yanase}, \citenamefont {Lv},\ and\ \citenamefont {Huang}}]{Chen2024prx}%
  \BibitemOpen
  \bibfield  {author} {\bibinfo {author} {\bibfnamefont {X.}~\bibnamefont {Chen}}, \bibinfo {author} {\bibfnamefont {L.}~\bibnamefont {Wang}}, \bibinfo {author} {\bibfnamefont {J.}~\bibnamefont {Ishizuka}}, \bibinfo {author} {\bibfnamefont {R.}~\bibnamefont {Zhang}}, \bibinfo {author} {\bibfnamefont {K.}~\bibnamefont {Nogaki}}, \bibinfo {author} {\bibfnamefont {Y.}~\bibnamefont {Cheng}}, \bibinfo {author} {\bibfnamefont {F.}~\bibnamefont {Yang}}, \bibinfo {author} {\bibfnamefont {Z.}~\bibnamefont {Chen}}, \bibinfo {author} {\bibfnamefont {F.}~\bibnamefont {Zhu}}, \bibinfo {author} {\bibfnamefont {Z.}~\bibnamefont {Liu}}, \bibinfo {author} {\bibfnamefont {J.}~\bibnamefont {Mei}}, \bibinfo {author} {\bibfnamefont {Y.}~\bibnamefont {Yanase}}, \bibinfo {author} {\bibfnamefont {B.}~\bibnamefont {Lv}},\ and\ \bibinfo {author} {\bibfnamefont {Y.}~\bibnamefont {Huang}},\ }\bibfield  {title} {\bibinfo {title} {Coexistence of near-${E}_{F}$ flat band and van hove singularity in a two-phase superconductor},\ }\href
  {https://doi.org/10.1103/PhysRevX.14.021048} {\bibfield  {journal} {\bibinfo  {journal} {Phys. Rev. X}\ }\textbf {\bibinfo {volume} {14}},\ \bibinfo {pages} {021048} (\bibinfo {year} {2024}{\natexlab{a}})}\BibitemShut {NoStop}%
\bibitem [{\citenamefont {Chen}\ \emph {et~al.}(2024{\natexlab{b}})\citenamefont {Chen}, \citenamefont {Liu}, \citenamefont {Wu}, \citenamefont {Zhang}, \citenamefont {Ye}, \citenamefont {Zhao}, \citenamefont {Song}, \citenamefont {Tian}, \citenamefont {Tan}, \citenamefont {Liu}, \citenamefont {Ye}, \citenamefont {Chen}, \citenamefont {Huang}, \citenamefont {Shen}, \citenamefont {Yuan}, \citenamefont {He}, \citenamefont {Duan},\ and\ \citenamefont {Meng}}]{Chen2024prb}%
  \BibitemOpen
  \bibfield  {author} {\bibinfo {author} {\bibfnamefont {B.}~\bibnamefont {Chen}}, \bibinfo {author} {\bibfnamefont {H.}~\bibnamefont {Liu}}, \bibinfo {author} {\bibfnamefont {Q.-Y.}\ \bibnamefont {Wu}}, \bibinfo {author} {\bibfnamefont {C.}~\bibnamefont {Zhang}}, \bibinfo {author} {\bibfnamefont {X.-Q.}\ \bibnamefont {Ye}}, \bibinfo {author} {\bibfnamefont {Y.-Z.}\ \bibnamefont {Zhao}}, \bibinfo {author} {\bibfnamefont {J.-J.}\ \bibnamefont {Song}}, \bibinfo {author} {\bibfnamefont {X.-Y.}\ \bibnamefont {Tian}}, \bibinfo {author} {\bibfnamefont {B.-L.}\ \bibnamefont {Tan}}, \bibinfo {author} {\bibfnamefont {Z.-T.}\ \bibnamefont {Liu}}, \bibinfo {author} {\bibfnamefont {M.}~\bibnamefont {Ye}}, \bibinfo {author} {\bibfnamefont {Z.-H.}\ \bibnamefont {Chen}}, \bibinfo {author} {\bibfnamefont {Y.-B.}\ \bibnamefont {Huang}}, \bibinfo {author} {\bibfnamefont {D.-W.}\ \bibnamefont {Shen}}, \bibinfo {author} {\bibfnamefont {Y.-H.}\ \bibnamefont {Yuan}}, \bibinfo {author} {\bibfnamefont {J.}~\bibnamefont {He}},
  \bibinfo {author} {\bibfnamefont {Y.-X.}\ \bibnamefont {Duan}},\ and\ \bibinfo {author} {\bibfnamefont {J.-Q.}\ \bibnamefont {Meng}},\ }\bibfield  {title} {\bibinfo {title} {Exploring possible fermi surface nesting and the nature of heavy quasiparticles in the spin-triplet superconductor candidate {${\mathrm{CeRh}}_{2}{\mathrm{As}}_{2}$}},\ }\href {https://doi.org/10.1103/PhysRevB.110.L041120} {\bibfield  {journal} {\bibinfo  {journal} {Phys. Rev. B}\ }\textbf {\bibinfo {volume} {110}},\ \bibinfo {pages} {L041120} (\bibinfo {year} {2024}{\natexlab{b}})}\BibitemShut {NoStop}%
\bibitem [{\citenamefont {Wu}\ \emph {et~al.}(2024)\citenamefont {Wu}, \citenamefont {Zhang}, \citenamefont {Ju}, \citenamefont {Hu}, \citenamefont {Huang}, \citenamefont {Zhang}, \citenamefont {Zhang}, \citenamefont {Zheng}, \citenamefont {Yang}, \citenamefont {Eljaouhari}, \citenamefont {Song}, \citenamefont {Plumb}, \citenamefont {Steglich}, \citenamefont {Shi}, \citenamefont {Zwicknagl}, \citenamefont {Cao}, \citenamefont {Yuan},\ and\ \citenamefont {Liu}}]{Wu2024cpl}%
  \BibitemOpen
  \bibfield  {author} {\bibinfo {author} {\bibfnamefont {Y.}~\bibnamefont {Wu}}, \bibinfo {author} {\bibfnamefont {Y.}~\bibnamefont {Zhang}}, \bibinfo {author} {\bibfnamefont {S.}~\bibnamefont {Ju}}, \bibinfo {author} {\bibfnamefont {Y.}~\bibnamefont {Hu}}, \bibinfo {author} {\bibfnamefont {Y.}~\bibnamefont {Huang}}, \bibinfo {author} {\bibfnamefont {Y.}~\bibnamefont {Zhang}}, \bibinfo {author} {\bibfnamefont {H.}~\bibnamefont {Zhang}}, \bibinfo {author} {\bibfnamefont {H.}~\bibnamefont {Zheng}}, \bibinfo {author} {\bibfnamefont {G.}~\bibnamefont {Yang}}, \bibinfo {author} {\bibfnamefont {E.-O.}\ \bibnamefont {Eljaouhari}}, \bibinfo {author} {\bibfnamefont {B.}~\bibnamefont {Song}}, \bibinfo {author} {\bibfnamefont {N.~C.}\ \bibnamefont {Plumb}}, \bibinfo {author} {\bibfnamefont {F.}~\bibnamefont {Steglich}}, \bibinfo {author} {\bibfnamefont {M.}~\bibnamefont {Shi}}, \bibinfo {author} {\bibfnamefont {G.}~\bibnamefont {Zwicknagl}}, \bibinfo {author} {\bibfnamefont {C.}~\bibnamefont {Cao}}, \bibinfo {author}
  {\bibfnamefont {H.}~\bibnamefont {Yuan}},\ and\ \bibinfo {author} {\bibfnamefont {Y.}~\bibnamefont {Liu}},\ }\bibfield  {title} {\bibinfo {title} {Fermi surface nesting with heavy quasiparticles in the locally noncentrosymmetric superconductor {CeRh$_2$As$_2$}},\ }\href {https://doi.org/10.1088/0256-307X/41/9/097403} {\bibfield  {journal} {\bibinfo  {journal} {Chinese Physics Letters}\ }\textbf {\bibinfo {volume} {41}},\ \bibinfo {pages} {097403} (\bibinfo {year} {2024})}\BibitemShut {NoStop}%
\bibitem [{\citenamefont {Pfeiffer}\ \emph {et~al.}(2024)\citenamefont {Pfeiffer}, \citenamefont {Semeniuk}, \citenamefont {Landaeta}, \citenamefont {Borth}, \citenamefont {Geibel}, \citenamefont {Nicklas}, \citenamefont {Brando}, \citenamefont {Khim},\ and\ \citenamefont {Hassinger}}]{Pfeiffer2024prl}%
  \BibitemOpen
  \bibfield  {author} {\bibinfo {author} {\bibfnamefont {M.}~\bibnamefont {Pfeiffer}}, \bibinfo {author} {\bibfnamefont {K.}~\bibnamefont {Semeniuk}}, \bibinfo {author} {\bibfnamefont {J.~F.}\ \bibnamefont {Landaeta}}, \bibinfo {author} {\bibfnamefont {R.}~\bibnamefont {Borth}}, \bibinfo {author} {\bibfnamefont {C.}~\bibnamefont {Geibel}}, \bibinfo {author} {\bibfnamefont {M.}~\bibnamefont {Nicklas}}, \bibinfo {author} {\bibfnamefont {M.}~\bibnamefont {Brando}}, \bibinfo {author} {\bibfnamefont {S.}~\bibnamefont {Khim}},\ and\ \bibinfo {author} {\bibfnamefont {E.}~\bibnamefont {Hassinger}},\ }\bibfield  {title} {\bibinfo {title} {Pressure-tuned quantum criticality in the locally noncentrosymmetric superconductor ${\mathrm{cerh}}_{2}{\mathrm{as}}_{2}$},\ }\href {https://doi.org/10.1103/PhysRevLett.133.126506} {\bibfield  {journal} {\bibinfo  {journal} {Phys. Rev. Lett.}\ }\textbf {\bibinfo {volume} {133}},\ \bibinfo {pages} {126506} (\bibinfo {year} {2024})}\BibitemShut {NoStop}%
\bibitem [{\citenamefont {Semeniuk}\ \emph {et~al.}(2024)\citenamefont {Semeniuk}, \citenamefont {Pfeiffer}, \citenamefont {Landaeta}, \citenamefont {Nicklas}, \citenamefont {Geibel}, \citenamefont {Brando}, \citenamefont {Khim},\ and\ \citenamefont {Hassinger}}]{Semeniuk2024prb}%
  \BibitemOpen
  \bibfield  {author} {\bibinfo {author} {\bibfnamefont {K.}~\bibnamefont {Semeniuk}}, \bibinfo {author} {\bibfnamefont {M.}~\bibnamefont {Pfeiffer}}, \bibinfo {author} {\bibfnamefont {J.~F.}\ \bibnamefont {Landaeta}}, \bibinfo {author} {\bibfnamefont {M.}~\bibnamefont {Nicklas}}, \bibinfo {author} {\bibfnamefont {C.}~\bibnamefont {Geibel}}, \bibinfo {author} {\bibfnamefont {M.}~\bibnamefont {Brando}}, \bibinfo {author} {\bibfnamefont {S.}~\bibnamefont {Khim}},\ and\ \bibinfo {author} {\bibfnamefont {E.}~\bibnamefont {Hassinger}},\ }\bibfield  {title} {\bibinfo {title} {Exposing the odd-parity superconductivity in ${\mathrm{cerh}}_{2}{\mathrm{as}}_{2}$ with hydrostatic pressure},\ }\href {https://doi.org/10.1103/PhysRevB.110.L100504} {\bibfield  {journal} {\bibinfo  {journal} {Phys. Rev. B}\ }\textbf {\bibinfo {volume} {110}},\ \bibinfo {pages} {L100504} (\bibinfo {year} {2024})}\BibitemShut {NoStop}%
\bibitem [{\citenamefont {Ogata}\ \emph {et~al.}(2024)\citenamefont {Ogata}, \citenamefont {Kitagawa}, \citenamefont {Kinjo}, \citenamefont {Ishida}, \citenamefont {Brando}, \citenamefont {Hassinger}, \citenamefont {Geibel},\ and\ \citenamefont {Khim}}]{Ogata2024prb}%
  \BibitemOpen
  \bibfield  {author} {\bibinfo {author} {\bibfnamefont {S.}~\bibnamefont {Ogata}}, \bibinfo {author} {\bibfnamefont {S.}~\bibnamefont {Kitagawa}}, \bibinfo {author} {\bibfnamefont {K.}~\bibnamefont {Kinjo}}, \bibinfo {author} {\bibfnamefont {K.}~\bibnamefont {Ishida}}, \bibinfo {author} {\bibfnamefont {M.}~\bibnamefont {Brando}}, \bibinfo {author} {\bibfnamefont {E.}~\bibnamefont {Hassinger}}, \bibinfo {author} {\bibfnamefont {C.}~\bibnamefont {Geibel}},\ and\ \bibinfo {author} {\bibfnamefont {S.}~\bibnamefont {Khim}},\ }\bibfield  {title} {\bibinfo {title} {Appearance of $c$-axis magnetic moment in odd-parity antiferromagnetic state in ${\mathrm{cerh}}_{2}{\mathrm{as}}_{2}$ revealed by {$^{75}\mathrm{As}$-NMR}},\ }\href {https://doi.org/10.1103/PhysRevB.110.214509} {\bibfield  {journal} {\bibinfo  {journal} {Phys. Rev. B}\ }\textbf {\bibinfo {volume} {110}},\ \bibinfo {pages} {214509} (\bibinfo {year} {2024})}\BibitemShut {NoStop}%
\bibitem [{\citenamefont {Chen}\ \emph {et~al.}(2024{\natexlab{c}})\citenamefont {Chen}, \citenamefont {Siddiquee}, \citenamefont {Xu}, \citenamefont {Rehfuss}, \citenamefont {Gao}, \citenamefont {Lygouras}, \citenamefont {Drouin}, \citenamefont {Morano}, \citenamefont {Avers}, \citenamefont {Schmitt}, \citenamefont {Podlesnyak}, \citenamefont {Paglione}, \citenamefont {Ran}, \citenamefont {Song},\ and\ \citenamefont {Broholm}}]{Chen2024prl}%
  \BibitemOpen
  \bibfield  {author} {\bibinfo {author} {\bibfnamefont {T.}~\bibnamefont {Chen}}, \bibinfo {author} {\bibfnamefont {H.}~\bibnamefont {Siddiquee}}, \bibinfo {author} {\bibfnamefont {Q.}~\bibnamefont {Xu}}, \bibinfo {author} {\bibfnamefont {Z.}~\bibnamefont {Rehfuss}}, \bibinfo {author} {\bibfnamefont {S.}~\bibnamefont {Gao}}, \bibinfo {author} {\bibfnamefont {C.}~\bibnamefont {Lygouras}}, \bibinfo {author} {\bibfnamefont {J.}~\bibnamefont {Drouin}}, \bibinfo {author} {\bibfnamefont {V.}~\bibnamefont {Morano}}, \bibinfo {author} {\bibfnamefont {K.~E.}\ \bibnamefont {Avers}}, \bibinfo {author} {\bibfnamefont {C.~J.}\ \bibnamefont {Schmitt}}, \bibinfo {author} {\bibfnamefont {A.}~\bibnamefont {Podlesnyak}}, \bibinfo {author} {\bibfnamefont {J.}~\bibnamefont {Paglione}}, \bibinfo {author} {\bibfnamefont {S.}~\bibnamefont {Ran}}, \bibinfo {author} {\bibfnamefont {Y.}~\bibnamefont {Song}},\ and\ \bibinfo {author} {\bibfnamefont {C.}~\bibnamefont {Broholm}},\ }\bibfield  {title} {\bibinfo {title}
  {Quasi-two-dimensional antiferromagnetic spin fluctuations in the spin-triplet superconductor candidate ${\mathrm{cerh}}_{2}{\mathrm{as}}_{2}$},\ }\href {https://doi.org/10.1103/PhysRevLett.133.266505} {\bibfield  {journal} {\bibinfo  {journal} {Phys. Rev. Lett.}\ }\textbf {\bibinfo {volume} {133}},\ \bibinfo {pages} {266505} (\bibinfo {year} {2024}{\natexlab{c}})}\BibitemShut {NoStop}%
\bibitem [{\citenamefont {Nogaki}\ \emph {et~al.}(2021)\citenamefont {Nogaki}, \citenamefont {Daido}, \citenamefont {Ishizuka},\ and\ \citenamefont {Yanase}}]{Nogaki2021prr}%
  \BibitemOpen
  \bibfield  {author} {\bibinfo {author} {\bibfnamefont {K.}~\bibnamefont {Nogaki}}, \bibinfo {author} {\bibfnamefont {A.}~\bibnamefont {Daido}}, \bibinfo {author} {\bibfnamefont {J.}~\bibnamefont {Ishizuka}},\ and\ \bibinfo {author} {\bibfnamefont {Y.}~\bibnamefont {Yanase}},\ }\bibfield  {title} {\bibinfo {title} {Topological crystalline superconductivity in locally noncentrosymmetric ${\mathrm{cerh}}_{2}{\mathrm{as}}_{2}$},\ }\href {https://doi.org/10.1103/PhysRevResearch.3.L032071} {\bibfield  {journal} {\bibinfo  {journal} {Phys. Rev. Res.}\ }\textbf {\bibinfo {volume} {3}},\ \bibinfo {pages} {L032071} (\bibinfo {year} {2021})}\BibitemShut {NoStop}%
\bibitem [{\citenamefont {Ishizuka}\ \emph {et~al.}(2024)\citenamefont {Ishizuka}, \citenamefont {Nogaki}, \citenamefont {Sigrist},\ and\ \citenamefont {Yanase}}]{Ishizuka2024prb}%
  \BibitemOpen
  \bibfield  {author} {\bibinfo {author} {\bibfnamefont {J.}~\bibnamefont {Ishizuka}}, \bibinfo {author} {\bibfnamefont {K.}~\bibnamefont {Nogaki}}, \bibinfo {author} {\bibfnamefont {M.}~\bibnamefont {Sigrist}},\ and\ \bibinfo {author} {\bibfnamefont {Y.}~\bibnamefont {Yanase}},\ }\bibfield  {title} {\bibinfo {title} {Correlation-induced fermi surface evolution and topological crystalline superconductivity in {${\mathrm{CeRh}}_{2}{\mathrm{As}}_{2}$}},\ }\href {https://doi.org/10.1103/PhysRevB.110.L140505} {\bibfield  {journal} {\bibinfo  {journal} {Phys. Rev. B}\ }\textbf {\bibinfo {volume} {110}},\ \bibinfo {pages} {L140505} (\bibinfo {year} {2024})}\BibitemShut {NoStop}%
\bibitem [{\citenamefont {Nogaki}\ and\ \citenamefont {Yanase}(2022)}]{Nogaki2022prb}%
  \BibitemOpen
  \bibfield  {author} {\bibinfo {author} {\bibfnamefont {K.}~\bibnamefont {Nogaki}}\ and\ \bibinfo {author} {\bibfnamefont {Y.}~\bibnamefont {Yanase}},\ }\bibfield  {title} {\bibinfo {title} {Even-odd parity transition in strongly correlated locally noncentrosymmetric superconductors: Application to ${\mathrm{cerh}}_{2}{\mathrm{as}}_{2}$},\ }\href {https://doi.org/10.1103/PhysRevB.106.L100504} {\bibfield  {journal} {\bibinfo  {journal} {Phys. Rev. B}\ }\textbf {\bibinfo {volume} {106}},\ \bibinfo {pages} {L100504} (\bibinfo {year} {2022})}\BibitemShut {NoStop}%
\bibitem [{\citenamefont {Lee}\ \emph {et~al.}(2025)\citenamefont {Lee}, \citenamefont {Agterberg},\ and\ \citenamefont {Brydon}}]{Lee2025prl}%
  \BibitemOpen
  \bibfield  {author} {\bibinfo {author} {\bibfnamefont {C.}~\bibnamefont {Lee}}, \bibinfo {author} {\bibfnamefont {D.~F.}\ \bibnamefont {Agterberg}},\ and\ \bibinfo {author} {\bibfnamefont {P.~M.~R.}\ \bibnamefont {Brydon}},\ }\bibfield  {title} {\bibinfo {title} {Unified picture of superconductivity and magnetism in ${\mathrm{cerh}}_{2}{\mathrm{as}}_{2}$},\ }\href {https://doi.org/10.1103/dt7w-hh9f} {\bibfield  {journal} {\bibinfo  {journal} {Phys. Rev. Lett.}\ }\textbf {\bibinfo {volume} {135}},\ \bibinfo {pages} {026003} (\bibinfo {year} {2025})}\BibitemShut {NoStop}%
\bibitem [{\citenamefont {Resta}(2011)}]{Resta2011}%
  \BibitemOpen
  \bibfield  {author} {\bibinfo {author} {\bibfnamefont {R.}~\bibnamefont {Resta}},\ }\bibfield  {title} {\bibinfo {title} {The insulating state of matter: a geometrical theory},\ }\href {https://link.springer.com/article/10.1140/epjb/e2010-10874-4} {\bibfield  {journal} {\bibinfo  {journal} {Eur. Phys. J. B}\ }\textbf {\bibinfo {volume} {79}},\ \bibinfo {pages} {121} (\bibinfo {year} {2011})}\BibitemShut {NoStop}%
\bibitem [{\citenamefont {T{\"o}rm{\"a}}\ \emph {et~al.}(2022)\citenamefont {T{\"o}rm{\"a}}, \citenamefont {Peotta},\ and\ \citenamefont {Bernevig}}]{Torma2022nrp}%
  \BibitemOpen
  \bibfield  {author} {\bibinfo {author} {\bibfnamefont {P.}~\bibnamefont {T{\"o}rm{\"a}}}, \bibinfo {author} {\bibfnamefont {S.}~\bibnamefont {Peotta}},\ and\ \bibinfo {author} {\bibfnamefont {B.~A.}\ \bibnamefont {Bernevig}},\ }\bibfield  {title} {\bibinfo {title} {Superconductivity, superfluidity and quantum geometry in twisted multilayer systems},\ }\href {https://doi.org/10.1038/s42254-022-00466-y} {\bibfield  {journal} {\bibinfo  {journal} {Nature Reviews Physics}\ }\textbf {\bibinfo {volume} {4}},\ \bibinfo {pages} {528} (\bibinfo {year} {2022})}\BibitemShut {NoStop}%
\bibitem [{\citenamefont {Liu}\ \emph {et~al.}(2024)\citenamefont {Liu}, \citenamefont {Qiang}, \citenamefont {Lu},\ and\ \citenamefont {Xie}}]{Liu2024nsr}%
  \BibitemOpen
  \bibfield  {author} {\bibinfo {author} {\bibfnamefont {T.}~\bibnamefont {Liu}}, \bibinfo {author} {\bibfnamefont {X.-B.}\ \bibnamefont {Qiang}}, \bibinfo {author} {\bibfnamefont {H.-Z.}\ \bibnamefont {Lu}},\ and\ \bibinfo {author} {\bibfnamefont {X.~C.}\ \bibnamefont {Xie}},\ }\bibfield  {title} {\bibinfo {title} {Quantum geometry in condensed matter},\ }\href {https://doi.org/10.1093/nsr/nwae334} {\bibfield  {journal} {\bibinfo  {journal} {National Science Review}\ }\textbf {\bibinfo {volume} {12}},\ \bibinfo {pages} {nwae334} (\bibinfo {year} {2024})}\BibitemShut {NoStop}%
\bibitem [{\citenamefont {Gao}\ \emph {et~al.}(2014)\citenamefont {Gao}, \citenamefont {Yang},\ and\ \citenamefont {Niu}}]{Gao2014}%
  \BibitemOpen
  \bibfield  {author} {\bibinfo {author} {\bibfnamefont {Y.}~\bibnamefont {Gao}}, \bibinfo {author} {\bibfnamefont {S.~A.}\ \bibnamefont {Yang}},\ and\ \bibinfo {author} {\bibfnamefont {Q.}~\bibnamefont {Niu}},\ }\bibfield  {title} {\bibinfo {title} {Field induced positional shift of bloch electrons and its dynamical implications},\ }\href {https://journals.aps.org/prl/abstract/10.1103/PhysRevLett.112.166601} {\bibfield  {journal} {\bibinfo  {journal} {Phys. Rev. Lett.}\ }\textbf {\bibinfo {volume} {112}},\ \bibinfo {pages} {166601} (\bibinfo {year} {2014})}\BibitemShut {NoStop}%
\bibitem [{\citenamefont {Sodemann}\ and\ \citenamefont {Fu}(2015)}]{Sodemann2015prl}%
  \BibitemOpen
  \bibfield  {author} {\bibinfo {author} {\bibfnamefont {I.}~\bibnamefont {Sodemann}}\ and\ \bibinfo {author} {\bibfnamefont {L.}~\bibnamefont {Fu}},\ }\bibfield  {title} {\bibinfo {title} {Quantum nonlinear hall effect induced by berry curvature dipole in time-reversal invariant materials},\ }\href {https://doi.org/10.1103/PhysRevLett.115.216806} {\bibfield  {journal} {\bibinfo  {journal} {Phys. Rev. Lett.}\ }\textbf {\bibinfo {volume} {115}},\ \bibinfo {pages} {216806} (\bibinfo {year} {2015})}\BibitemShut {NoStop}%
\bibitem [{\citenamefont {Du}\ \emph {et~al.}(2018)\citenamefont {Du}, \citenamefont {Wang}, \citenamefont {Lu},\ and\ \citenamefont {Xie}}]{Du2018prl}%
  \BibitemOpen
  \bibfield  {author} {\bibinfo {author} {\bibfnamefont {Z.~Z.}\ \bibnamefont {Du}}, \bibinfo {author} {\bibfnamefont {C.~M.}\ \bibnamefont {Wang}}, \bibinfo {author} {\bibfnamefont {H.-Z.}\ \bibnamefont {Lu}},\ and\ \bibinfo {author} {\bibfnamefont {X.~C.}\ \bibnamefont {Xie}},\ }\bibfield  {title} {\bibinfo {title} {Band signatures for strong nonlinear hall effect in bilayer ${\mathrm{wte}}_{2}$},\ }\href {https://doi.org/10.1103/PhysRevLett.121.266601} {\bibfield  {journal} {\bibinfo  {journal} {Phys. Rev. Lett.}\ }\textbf {\bibinfo {volume} {121}},\ \bibinfo {pages} {266601} (\bibinfo {year} {2018})}\BibitemShut {NoStop}%
\bibitem [{\citenamefont {Du}\ \emph {et~al.}(2019)\citenamefont {Du}, \citenamefont {Wang}, \citenamefont {Li}, \citenamefont {Lu},\ and\ \citenamefont {Xie}}]{Du2019nc}%
  \BibitemOpen
  \bibfield  {author} {\bibinfo {author} {\bibfnamefont {Z.~Z.}\ \bibnamefont {Du}}, \bibinfo {author} {\bibfnamefont {C.~M.}\ \bibnamefont {Wang}}, \bibinfo {author} {\bibfnamefont {S.}~\bibnamefont {Li}}, \bibinfo {author} {\bibfnamefont {H.-Z.}\ \bibnamefont {Lu}},\ and\ \bibinfo {author} {\bibfnamefont {X.~C.}\ \bibnamefont {Xie}},\ }\bibfield  {title} {\bibinfo {title} {Disorder-induced nonlinear hall effect with time-reversal symmetry},\ }\href {https://doi.org/10.1038/s41467-019-10941-3} {\bibfield  {journal} {\bibinfo  {journal} {Nature Communications}\ }\textbf {\bibinfo {volume} {10}},\ \bibinfo {pages} {3047} (\bibinfo {year} {2019})}\BibitemShut {NoStop}%
\bibitem [{\citenamefont {Ma}\ \emph {et~al.}(2019)\citenamefont {Ma}, \citenamefont {Xu}, \citenamefont {Shen}, \citenamefont {MacNeill}, \citenamefont {Fatemi}, \citenamefont {Chang}, \citenamefont {Mier~Valdivia}, \citenamefont {Wu}, \citenamefont {Du}, \citenamefont {Hsu}, \citenamefont {Fang}, \citenamefont {Gibson}, \citenamefont {Watanabe}, \citenamefont {Taniguchi}, \citenamefont {Cava}, \citenamefont {Kaxiras}, \citenamefont {Lu}, \citenamefont {Lin}, \citenamefont {Fu}, \citenamefont {Gedik},\ and\ \citenamefont {Jarillo-Herrero}}]{Ma2019nature}%
  \BibitemOpen
  \bibfield  {author} {\bibinfo {author} {\bibfnamefont {Q.}~\bibnamefont {Ma}}, \bibinfo {author} {\bibfnamefont {S.-Y.}\ \bibnamefont {Xu}}, \bibinfo {author} {\bibfnamefont {H.}~\bibnamefont {Shen}}, \bibinfo {author} {\bibfnamefont {D.}~\bibnamefont {MacNeill}}, \bibinfo {author} {\bibfnamefont {V.}~\bibnamefont {Fatemi}}, \bibinfo {author} {\bibfnamefont {T.-R.}\ \bibnamefont {Chang}}, \bibinfo {author} {\bibfnamefont {A.~M.}\ \bibnamefont {Mier~Valdivia}}, \bibinfo {author} {\bibfnamefont {S.}~\bibnamefont {Wu}}, \bibinfo {author} {\bibfnamefont {Z.}~\bibnamefont {Du}}, \bibinfo {author} {\bibfnamefont {C.-H.}\ \bibnamefont {Hsu}}, \bibinfo {author} {\bibfnamefont {S.}~\bibnamefont {Fang}}, \bibinfo {author} {\bibfnamefont {Q.~D.}\ \bibnamefont {Gibson}}, \bibinfo {author} {\bibfnamefont {K.}~\bibnamefont {Watanabe}}, \bibinfo {author} {\bibfnamefont {T.}~\bibnamefont {Taniguchi}}, \bibinfo {author} {\bibfnamefont {R.~J.}\ \bibnamefont {Cava}}, \bibinfo {author} {\bibfnamefont {E.}~\bibnamefont
  {Kaxiras}}, \bibinfo {author} {\bibfnamefont {H.-Z.}\ \bibnamefont {Lu}}, \bibinfo {author} {\bibfnamefont {H.}~\bibnamefont {Lin}}, \bibinfo {author} {\bibfnamefont {L.}~\bibnamefont {Fu}}, \bibinfo {author} {\bibfnamefont {N.}~\bibnamefont {Gedik}},\ and\ \bibinfo {author} {\bibfnamefont {P.}~\bibnamefont {Jarillo-Herrero}},\ }\bibfield  {title} {\bibinfo {title} {Observation of the nonlinear hall effect under time-reversal-symmetric conditions},\ }\href {https://doi.org/10.1038/s41586-018-0807-6} {\bibfield  {journal} {\bibinfo  {journal} {Nature}\ }\textbf {\bibinfo {volume} {565}},\ \bibinfo {pages} {337} (\bibinfo {year} {2019})}\BibitemShut {NoStop}%
\bibitem [{\citenamefont {Wang}\ \emph {et~al.}(2023)\citenamefont {Wang}, \citenamefont {Kaplan}, \citenamefont {Zhang}, \citenamefont {Holder}, \citenamefont {Cao}, \citenamefont {Wang}, \citenamefont {Zhou}, \citenamefont {Zhou}, \citenamefont {Jiang}, \citenamefont {Zhang}, \citenamefont {Ru}, \citenamefont {Cai}, \citenamefont {Watanabe}, \citenamefont {Taniguchi}, \citenamefont {Yan},\ and\ \citenamefont {Gao}}]{Wang2023}%
  \BibitemOpen
  \bibfield  {author} {\bibinfo {author} {\bibfnamefont {N.}~\bibnamefont {Wang}}, \bibinfo {author} {\bibfnamefont {D.}~\bibnamefont {Kaplan}}, \bibinfo {author} {\bibfnamefont {Z.}~\bibnamefont {Zhang}}, \bibinfo {author} {\bibfnamefont {T.}~\bibnamefont {Holder}}, \bibinfo {author} {\bibfnamefont {N.}~\bibnamefont {Cao}}, \bibinfo {author} {\bibfnamefont {A.}~\bibnamefont {Wang}}, \bibinfo {author} {\bibfnamefont {X.}~\bibnamefont {Zhou}}, \bibinfo {author} {\bibfnamefont {F.}~\bibnamefont {Zhou}}, \bibinfo {author} {\bibfnamefont {Z.}~\bibnamefont {Jiang}}, \bibinfo {author} {\bibfnamefont {C.}~\bibnamefont {Zhang}}, \bibinfo {author} {\bibfnamefont {S.}~\bibnamefont {Ru}}, \bibinfo {author} {\bibfnamefont {H.}~\bibnamefont {Cai}}, \bibinfo {author} {\bibfnamefont {K.}~\bibnamefont {Watanabe}}, \bibinfo {author} {\bibfnamefont {T.}~\bibnamefont {Taniguchi}}, \bibinfo {author} {\bibfnamefont {B.}~\bibnamefont {Yan}},\ and\ \bibinfo {author} {\bibfnamefont {W.}~\bibnamefont {Gao}},\ }\bibfield  {title}
  {\bibinfo {title} {Quantum-metric-induced nonlinear transport in a topological antiferromagnet},\ }\href {https://doi.org/10.1038/s41586-023-06363-3} {\bibfield  {journal} {\bibinfo  {journal} {Nature}\ }\textbf {\bibinfo {volume} {621}},\ \bibinfo {pages} {487} (\bibinfo {year} {2023})}\BibitemShut {NoStop}%
\bibitem [{\citenamefont {Gao}\ \emph {et~al.}(2023)\citenamefont {Gao}, \citenamefont {Liu}, \citenamefont {Qiu}, \citenamefont {Ghosh}, \citenamefont {Trevisan}, \citenamefont {Onishi}, \citenamefont {Hu}, \citenamefont {Qian}, \citenamefont {Tien}, \citenamefont {Chen}, \citenamefont {Huang}, \citenamefont {Bérubé}, \citenamefont {Li}, \citenamefont {Tzschaschel}, \citenamefont {Dinh}, \citenamefont {Sun}, \citenamefont {Ho}, \citenamefont {Lien}, \citenamefont {Singh}, \citenamefont {Watanabe}, \citenamefont {Taniguchi}, \citenamefont {Bell}, \citenamefont {Lin}, \citenamefont {Chang}, \citenamefont {Du}, \citenamefont {Bansil}, \citenamefont {Fu}, \citenamefont {Ni}, \citenamefont {Orth}, \citenamefont {Ma},\ and\ \citenamefont {Xu}}]{Gao2023}%
  \BibitemOpen
  \bibfield  {author} {\bibinfo {author} {\bibfnamefont {A.}~\bibnamefont {Gao}}, \bibinfo {author} {\bibfnamefont {Y.-F.}\ \bibnamefont {Liu}}, \bibinfo {author} {\bibfnamefont {J.-X.}\ \bibnamefont {Qiu}}, \bibinfo {author} {\bibfnamefont {B.}~\bibnamefont {Ghosh}}, \bibinfo {author} {\bibfnamefont {T.~V.}\ \bibnamefont {Trevisan}}, \bibinfo {author} {\bibfnamefont {Y.}~\bibnamefont {Onishi}}, \bibinfo {author} {\bibfnamefont {C.}~\bibnamefont {Hu}}, \bibinfo {author} {\bibfnamefont {T.}~\bibnamefont {Qian}}, \bibinfo {author} {\bibfnamefont {H.-J.}\ \bibnamefont {Tien}}, \bibinfo {author} {\bibfnamefont {S.-W.}\ \bibnamefont {Chen}}, \bibinfo {author} {\bibfnamefont {M.}~\bibnamefont {Huang}}, \bibinfo {author} {\bibfnamefont {D.}~\bibnamefont {Bérubé}}, \bibinfo {author} {\bibfnamefont {H.}~\bibnamefont {Li}}, \bibinfo {author} {\bibfnamefont {C.}~\bibnamefont {Tzschaschel}}, \bibinfo {author} {\bibfnamefont {T.}~\bibnamefont {Dinh}}, \bibinfo {author} {\bibfnamefont {Z.}~\bibnamefont {Sun}}, \bibinfo
  {author} {\bibfnamefont {S.-C.}\ \bibnamefont {Ho}}, \bibinfo {author} {\bibfnamefont {S.-W.}\ \bibnamefont {Lien}}, \bibinfo {author} {\bibfnamefont {B.}~\bibnamefont {Singh}}, \bibinfo {author} {\bibfnamefont {K.}~\bibnamefont {Watanabe}}, \bibinfo {author} {\bibfnamefont {T.}~\bibnamefont {Taniguchi}}, \bibinfo {author} {\bibfnamefont {D.~C.}\ \bibnamefont {Bell}}, \bibinfo {author} {\bibfnamefont {H.}~\bibnamefont {Lin}}, \bibinfo {author} {\bibfnamefont {T.-R.}\ \bibnamefont {Chang}}, \bibinfo {author} {\bibfnamefont {C.~R.}\ \bibnamefont {Du}}, \bibinfo {author} {\bibfnamefont {A.}~\bibnamefont {Bansil}}, \bibinfo {author} {\bibfnamefont {L.}~\bibnamefont {Fu}}, \bibinfo {author} {\bibfnamefont {N.}~\bibnamefont {Ni}}, \bibinfo {author} {\bibfnamefont {P.~P.}\ \bibnamefont {Orth}}, \bibinfo {author} {\bibfnamefont {Q.}~\bibnamefont {Ma}},\ and\ \bibinfo {author} {\bibfnamefont {S.-Y.}\ \bibnamefont {Xu}},\ }\bibfield  {title} {\bibinfo {title} {Quantum metric nonlinear hall effect in a topological
  antiferromagnetic heterostructure},\ }\href {https://doi.org/10.1126/science.adf1506} {\bibfield  {journal} {\bibinfo  {journal} {Science}\ }\textbf {\bibinfo {volume} {381}},\ \bibinfo {pages} {181} (\bibinfo {year} {2023})}\BibitemShut {NoStop}%
\bibitem [{\citenamefont {Han}\ \emph {et~al.}(2024)\citenamefont {Han}, \citenamefont {Uchimura}, \citenamefont {Araki}, \citenamefont {Yoon}, \citenamefont {Takeuchi}, \citenamefont {Yamane}, \citenamefont {Kanai}, \citenamefont {Ieda}, \citenamefont {Ohno},\ and\ \citenamefont {Fukami}}]{Han2024}%
  \BibitemOpen
  \bibfield  {author} {\bibinfo {author} {\bibfnamefont {J.}~\bibnamefont {Han}}, \bibinfo {author} {\bibfnamefont {T.}~\bibnamefont {Uchimura}}, \bibinfo {author} {\bibfnamefont {Y.}~\bibnamefont {Araki}}, \bibinfo {author} {\bibfnamefont {J.-Y.}\ \bibnamefont {Yoon}}, \bibinfo {author} {\bibfnamefont {Y.}~\bibnamefont {Takeuchi}}, \bibinfo {author} {\bibfnamefont {Y.}~\bibnamefont {Yamane}}, \bibinfo {author} {\bibfnamefont {S.}~\bibnamefont {Kanai}}, \bibinfo {author} {\bibfnamefont {J.}~\bibnamefont {Ieda}}, \bibinfo {author} {\bibfnamefont {H.}~\bibnamefont {Ohno}},\ and\ \bibinfo {author} {\bibfnamefont {S.}~\bibnamefont {Fukami}},\ }\bibfield  {title} {\bibinfo {title} {Room-temperature flexible manipulation of the quantum-metric structure in a topological chiral antiferromagnet},\ }\href {https://doi.org/10.1038/s41567-024-02476-2} {\bibfield  {journal} {\bibinfo  {journal} {Nature Physics}\ }\textbf {\bibinfo {volume} {20}},\ \bibinfo {pages} {1110} (\bibinfo {year} {2024})}\BibitemShut {NoStop}%
\bibitem [{\citenamefont {Orenstein}\ \emph {et~al.}(2021)\citenamefont {Orenstein}, \citenamefont {Moore}, \citenamefont {Morimoto}, \citenamefont {Torchinsky}, \citenamefont {Harter},\ and\ \citenamefont {Hsieh}}]{annualrevOrenstein}%
  \BibitemOpen
  \bibfield  {author} {\bibinfo {author} {\bibfnamefont {J.}~\bibnamefont {Orenstein}}, \bibinfo {author} {\bibfnamefont {J.}~\bibnamefont {Moore}}, \bibinfo {author} {\bibfnamefont {T.}~\bibnamefont {Morimoto}}, \bibinfo {author} {\bibfnamefont {D.}~\bibnamefont {Torchinsky}}, \bibinfo {author} {\bibfnamefont {J.}~\bibnamefont {Harter}},\ and\ \bibinfo {author} {\bibfnamefont {D.}~\bibnamefont {Hsieh}},\ }\bibfield  {title} {\bibinfo {title} {Topology and symmetry of quantum materials via nonlinear optical responses},\ }\href {https://doi.org/https://doi.org/10.1146/annurev-conmatphys-031218-013712} {\bibfield  {journal} {\bibinfo  {journal} {Annual Review of Condensed Matter Physics}\ }\textbf {\bibinfo {volume} {12}},\ \bibinfo {pages} {247} (\bibinfo {year} {2021})}\BibitemShut {NoStop}%
\bibitem [{\citenamefont {Morimoto}\ \emph {et~al.}(2023)\citenamefont {Morimoto}, \citenamefont {Kitamura},\ and\ \citenamefont {Nagaosa}}]{Morimoto_JPSJreview}%
  \BibitemOpen
  \bibfield  {author} {\bibinfo {author} {\bibfnamefont {T.}~\bibnamefont {Morimoto}}, \bibinfo {author} {\bibfnamefont {S.}~\bibnamefont {Kitamura}},\ and\ \bibinfo {author} {\bibfnamefont {N.}~\bibnamefont {Nagaosa}},\ }\bibfield  {title} {\bibinfo {title} {Geometric aspects of nonlinear and nonequilibrium phenomena},\ }\href {https://doi.org/10.7566/JPSJ.92.072001} {\bibfield  {journal} {\bibinfo  {journal} {J. Phys. Soc. Jpn.}\ }\textbf {\bibinfo {volume} {92}},\ \bibinfo {pages} {072001} (\bibinfo {year} {2023})}\BibitemShut {NoStop}%
\bibitem [{\citenamefont {Nagaosa}\ and\ \citenamefont {Yanase}(2024)}]{Nagaosa-Yanase}%
  \BibitemOpen
  \bibfield  {author} {\bibinfo {author} {\bibfnamefont {N.}~\bibnamefont {Nagaosa}}\ and\ \bibinfo {author} {\bibfnamefont {Y.}~\bibnamefont {Yanase}},\ }\bibfield  {title} {\bibinfo {title} {Nonreciprocal transport and optical phenomena in quantum materials},\ }\href {https://doi.org/https://doi.org/10.1146/annurev-conmatphys-032822-033734} {\bibfield  {journal} {\bibinfo  {journal} {Annu. Rev. Condens. Matter Phys.}\ }\textbf {\bibinfo {volume} {15}},\ \bibinfo {pages} {63} (\bibinfo {year} {2024})}\BibitemShut {NoStop}%
\bibitem [{\citenamefont {Peotta}\ and\ \citenamefont {T{\"o}rm{\"a}}(2015)}]{Peotta2015nc}%
  \BibitemOpen
  \bibfield  {author} {\bibinfo {author} {\bibfnamefont {S.}~\bibnamefont {Peotta}}\ and\ \bibinfo {author} {\bibfnamefont {P.}~\bibnamefont {T{\"o}rm{\"a}}},\ }\bibfield  {title} {\bibinfo {title} {Superfluidity in topologically nontrivial flat bands},\ }\href {https://doi.org/10.1038/ncomms9944} {\bibfield  {journal} {\bibinfo  {journal} {Nature Communications}\ }\textbf {\bibinfo {volume} {6}},\ \bibinfo {pages} {8944} (\bibinfo {year} {2015})}\BibitemShut {NoStop}%
\bibitem [{\citenamefont {Julku}\ \emph {et~al.}(2016)\citenamefont {Julku}, \citenamefont {Peotta}, \citenamefont {Vanhala}, \citenamefont {Kim},\ and\ \citenamefont {T\"orm\"a}}]{Julku2016prl}%
  \BibitemOpen
  \bibfield  {author} {\bibinfo {author} {\bibfnamefont {A.}~\bibnamefont {Julku}}, \bibinfo {author} {\bibfnamefont {S.}~\bibnamefont {Peotta}}, \bibinfo {author} {\bibfnamefont {T.~I.}\ \bibnamefont {Vanhala}}, \bibinfo {author} {\bibfnamefont {D.-H.}\ \bibnamefont {Kim}},\ and\ \bibinfo {author} {\bibfnamefont {P.}~\bibnamefont {T\"orm\"a}},\ }\bibfield  {title} {\bibinfo {title} {Geometric origin of superfluidity in the lieb-lattice flat band},\ }\href {https://doi.org/10.1103/PhysRevLett.117.045303} {\bibfield  {journal} {\bibinfo  {journal} {Phys. Rev. Lett.}\ }\textbf {\bibinfo {volume} {117}},\ \bibinfo {pages} {045303} (\bibinfo {year} {2016})}\BibitemShut {NoStop}%
\bibitem [{\citenamefont {Liang}\ \emph {et~al.}(2017)\citenamefont {Liang}, \citenamefont {Vanhala}, \citenamefont {Peotta}, \citenamefont {Siro}, \citenamefont {Harju},\ and\ \citenamefont {T\"orm\"a}}]{Liang2017prb}%
  \BibitemOpen
  \bibfield  {author} {\bibinfo {author} {\bibfnamefont {L.}~\bibnamefont {Liang}}, \bibinfo {author} {\bibfnamefont {T.~I.}\ \bibnamefont {Vanhala}}, \bibinfo {author} {\bibfnamefont {S.}~\bibnamefont {Peotta}}, \bibinfo {author} {\bibfnamefont {T.}~\bibnamefont {Siro}}, \bibinfo {author} {\bibfnamefont {A.}~\bibnamefont {Harju}},\ and\ \bibinfo {author} {\bibfnamefont {P.}~\bibnamefont {T\"orm\"a}},\ }\bibfield  {title} {\bibinfo {title} {Band geometry, berry curvature, and superfluid weight},\ }\href {https://doi.org/10.1103/PhysRevB.95.024515} {\bibfield  {journal} {\bibinfo  {journal} {Phys. Rev. B}\ }\textbf {\bibinfo {volume} {95}},\ \bibinfo {pages} {024515} (\bibinfo {year} {2017})}\BibitemShut {NoStop}%
\bibitem [{\citenamefont {Tian}\ \emph {et~al.}(2023)\citenamefont {Tian}, \citenamefont {Gao}, \citenamefont {Zhang}, \citenamefont {Che}, \citenamefont {Xu}, \citenamefont {Cheung}, \citenamefont {Watanabe}, \citenamefont {Taniguchi}, \citenamefont {Randeria}, \citenamefont {Zhang}, \citenamefont {Lau},\ and\ \citenamefont {Bockrath}}]{Tian2023}%
  \BibitemOpen
  \bibfield  {author} {\bibinfo {author} {\bibfnamefont {H.}~\bibnamefont {Tian}}, \bibinfo {author} {\bibfnamefont {X.}~\bibnamefont {Gao}}, \bibinfo {author} {\bibfnamefont {Y.}~\bibnamefont {Zhang}}, \bibinfo {author} {\bibfnamefont {S.}~\bibnamefont {Che}}, \bibinfo {author} {\bibfnamefont {T.}~\bibnamefont {Xu}}, \bibinfo {author} {\bibfnamefont {P.}~\bibnamefont {Cheung}}, \bibinfo {author} {\bibfnamefont {K.}~\bibnamefont {Watanabe}}, \bibinfo {author} {\bibfnamefont {T.}~\bibnamefont {Taniguchi}}, \bibinfo {author} {\bibfnamefont {M.}~\bibnamefont {Randeria}}, \bibinfo {author} {\bibfnamefont {F.}~\bibnamefont {Zhang}}, \bibinfo {author} {\bibfnamefont {C.~N.}\ \bibnamefont {Lau}},\ and\ \bibinfo {author} {\bibfnamefont {M.~W.}\ \bibnamefont {Bockrath}},\ }\bibfield  {title} {\bibinfo {title} {Evidence for dirac flat band superconductivity enabled by quantum geometry},\ }\href {https://www.nature.com/articles/s41586-022-05576-2} {\bibfield  {journal} {\bibinfo  {journal} {Nature}\ }\textbf {\bibinfo
  {volume} {614}},\ \bibinfo {pages} {440} (\bibinfo {year} {2023})}\BibitemShut {NoStop}%
\bibitem [{\citenamefont {Tanaka}\ \emph {et~al.}(2025)\citenamefont {Tanaka}, \citenamefont {Wang}, \citenamefont {Dinh}, \citenamefont {Rodan-Legrain}, \citenamefont {Zaman}, \citenamefont {Hays}, \citenamefont {Almanakly}, \citenamefont {Kannan}, \citenamefont {Kim}, \citenamefont {Niedzielski}, \citenamefont {Serniak}, \citenamefont {Schwartz}, \citenamefont {Watanabe}, \citenamefont {Taniguchi}, \citenamefont {Orlando}, \citenamefont {Gustavsson}, \citenamefont {Grover}, \citenamefont {Jarillo-Herrero},\ and\ \citenamefont {Oliver}}]{Tanaka2025-sy}%
  \BibitemOpen
  \bibfield  {author} {\bibinfo {author} {\bibfnamefont {M.}~\bibnamefont {Tanaka}}, \bibinfo {author} {\bibfnamefont {J.~{\^{I}}.-J.}\ \bibnamefont {Wang}}, \bibinfo {author} {\bibfnamefont {T.~H.}\ \bibnamefont {Dinh}}, \bibinfo {author} {\bibfnamefont {D.}~\bibnamefont {Rodan-Legrain}}, \bibinfo {author} {\bibfnamefont {S.}~\bibnamefont {Zaman}}, \bibinfo {author} {\bibfnamefont {M.}~\bibnamefont {Hays}}, \bibinfo {author} {\bibfnamefont {A.}~\bibnamefont {Almanakly}}, \bibinfo {author} {\bibfnamefont {B.}~\bibnamefont {Kannan}}, \bibinfo {author} {\bibfnamefont {D.~K.}\ \bibnamefont {Kim}}, \bibinfo {author} {\bibfnamefont {B.~M.}\ \bibnamefont {Niedzielski}}, \bibinfo {author} {\bibfnamefont {K.}~\bibnamefont {Serniak}}, \bibinfo {author} {\bibfnamefont {M.~E.}\ \bibnamefont {Schwartz}}, \bibinfo {author} {\bibfnamefont {K.}~\bibnamefont {Watanabe}}, \bibinfo {author} {\bibfnamefont {T.}~\bibnamefont {Taniguchi}}, \bibinfo {author} {\bibfnamefont {T.~P.}\ \bibnamefont {Orlando}}, \bibinfo {author}
  {\bibfnamefont {S.}~\bibnamefont {Gustavsson}}, \bibinfo {author} {\bibfnamefont {J.~A.}\ \bibnamefont {Grover}}, \bibinfo {author} {\bibfnamefont {P.}~\bibnamefont {Jarillo-Herrero}},\ and\ \bibinfo {author} {\bibfnamefont {W.~D.}\ \bibnamefont {Oliver}},\ }\bibfield  {title} {\bibinfo {title} {Superfluid stiffness of magic-angle twisted bilayer graphene},\ }\href {https://doi.org/10.1038/s41586-024-08494-7} {\bibfield  {journal} {\bibinfo  {journal} {Nature}\ }\textbf {\bibinfo {volume} {638}},\ \bibinfo {pages} {99} (\bibinfo {year} {2025})}\BibitemShut {NoStop}%
\bibitem [{\citenamefont {Banerjee}\ \emph {et~al.}(2025)\citenamefont {Banerjee}, \citenamefont {Hao}, \citenamefont {Kreidel}, \citenamefont {Ledwith}, \citenamefont {Phinney}, \citenamefont {Park}, \citenamefont {Zimmerman}, \citenamefont {Wesson}, \citenamefont {Watanabe}, \citenamefont {Taniguchi}, \citenamefont {Westervelt}, \citenamefont {Yacoby}, \citenamefont {Jarillo-Herrero}, \citenamefont {Volkov}, \citenamefont {Vishwanath}, \citenamefont {Fong},\ and\ \citenamefont {Kim}}]{Banerjee2025-ew}%
  \BibitemOpen
  \bibfield  {author} {\bibinfo {author} {\bibfnamefont {A.}~\bibnamefont {Banerjee}}, \bibinfo {author} {\bibfnamefont {Z.}~\bibnamefont {Hao}}, \bibinfo {author} {\bibfnamefont {M.}~\bibnamefont {Kreidel}}, \bibinfo {author} {\bibfnamefont {P.}~\bibnamefont {Ledwith}}, \bibinfo {author} {\bibfnamefont {I.}~\bibnamefont {Phinney}}, \bibinfo {author} {\bibfnamefont {J.~M.}\ \bibnamefont {Park}}, \bibinfo {author} {\bibfnamefont {A.}~\bibnamefont {Zimmerman}}, \bibinfo {author} {\bibfnamefont {M.~E.}\ \bibnamefont {Wesson}}, \bibinfo {author} {\bibfnamefont {K.}~\bibnamefont {Watanabe}}, \bibinfo {author} {\bibfnamefont {T.}~\bibnamefont {Taniguchi}}, \bibinfo {author} {\bibfnamefont {R.~M.}\ \bibnamefont {Westervelt}}, \bibinfo {author} {\bibfnamefont {A.}~\bibnamefont {Yacoby}}, \bibinfo {author} {\bibfnamefont {P.}~\bibnamefont {Jarillo-Herrero}}, \bibinfo {author} {\bibfnamefont {P.~A.}\ \bibnamefont {Volkov}}, \bibinfo {author} {\bibfnamefont {A.}~\bibnamefont {Vishwanath}}, \bibinfo {author}
  {\bibfnamefont {K.~C.}\ \bibnamefont {Fong}},\ and\ \bibinfo {author} {\bibfnamefont {P.}~\bibnamefont {Kim}},\ }\bibfield  {title} {\bibinfo {title} {Superfluid stiffness of twisted trilayer graphene superconductors},\ }\href {https://doi.org/10.1038/s41586-024-08444-3} {\bibfield  {journal} {\bibinfo  {journal} {Nature}\ }\textbf {\bibinfo {volume} {638}},\ \bibinfo {pages} {93} (\bibinfo {year} {2025})}\BibitemShut {NoStop}%
\bibitem [{\citenamefont {Tang}\ \emph {et~al.}(2011)\citenamefont {Tang}, \citenamefont {Mei},\ and\ \citenamefont {Wen}}]{Tang2011prl}%
  \BibitemOpen
  \bibfield  {author} {\bibinfo {author} {\bibfnamefont {E.}~\bibnamefont {Tang}}, \bibinfo {author} {\bibfnamefont {J.-W.}\ \bibnamefont {Mei}},\ and\ \bibinfo {author} {\bibfnamefont {X.-G.}\ \bibnamefont {Wen}},\ }\bibfield  {title} {\bibinfo {title} {High-temperature fractional quantum hall states},\ }\href {https://doi.org/10.1103/PhysRevLett.106.236802} {\bibfield  {journal} {\bibinfo  {journal} {Phys. Rev. Lett.}\ }\textbf {\bibinfo {volume} {106}},\ \bibinfo {pages} {236802} (\bibinfo {year} {2011})}\BibitemShut {NoStop}%
\bibitem [{\citenamefont {Sun}\ \emph {et~al.}(2011)\citenamefont {Sun}, \citenamefont {Gu}, \citenamefont {Katsura},\ and\ \citenamefont {Das~Sarma}}]{Kai2011prl}%
  \BibitemOpen
  \bibfield  {author} {\bibinfo {author} {\bibfnamefont {K.}~\bibnamefont {Sun}}, \bibinfo {author} {\bibfnamefont {Z.}~\bibnamefont {Gu}}, \bibinfo {author} {\bibfnamefont {H.}~\bibnamefont {Katsura}},\ and\ \bibinfo {author} {\bibfnamefont {S.}~\bibnamefont {Das~Sarma}},\ }\bibfield  {title} {\bibinfo {title} {Nearly flatbands with nontrivial topology},\ }\href {https://doi.org/10.1103/PhysRevLett.106.236803} {\bibfield  {journal} {\bibinfo  {journal} {Phys. Rev. Lett.}\ }\textbf {\bibinfo {volume} {106}},\ \bibinfo {pages} {236803} (\bibinfo {year} {2011})}\BibitemShut {NoStop}%
\bibitem [{\citenamefont {Neupert}\ \emph {et~al.}(2011)\citenamefont {Neupert}, \citenamefont {Santos}, \citenamefont {Chamon},\ and\ \citenamefont {Mudry}}]{Neupert2011prl}%
  \BibitemOpen
  \bibfield  {author} {\bibinfo {author} {\bibfnamefont {T.}~\bibnamefont {Neupert}}, \bibinfo {author} {\bibfnamefont {L.}~\bibnamefont {Santos}}, \bibinfo {author} {\bibfnamefont {C.}~\bibnamefont {Chamon}},\ and\ \bibinfo {author} {\bibfnamefont {C.}~\bibnamefont {Mudry}},\ }\bibfield  {title} {\bibinfo {title} {Fractional quantum hall states at zero magnetic field},\ }\href {https://doi.org/10.1103/PhysRevLett.106.236804} {\bibfield  {journal} {\bibinfo  {journal} {Phys. Rev. Lett.}\ }\textbf {\bibinfo {volume} {106}},\ \bibinfo {pages} {236804} (\bibinfo {year} {2011})}\BibitemShut {NoStop}%
\bibitem [{\citenamefont {Kitamura}\ \emph {et~al.}(2024)\citenamefont {Kitamura}, \citenamefont {Daido},\ and\ \citenamefont {Yanase}}]{Kitamura2024prl}%
  \BibitemOpen
  \bibfield  {author} {\bibinfo {author} {\bibfnamefont {T.}~\bibnamefont {Kitamura}}, \bibinfo {author} {\bibfnamefont {A.}~\bibnamefont {Daido}},\ and\ \bibinfo {author} {\bibfnamefont {Y.}~\bibnamefont {Yanase}},\ }\bibfield  {title} {\bibinfo {title} {Spin-triplet superconductivity from quantum-geometry-induced ferromagnetic fluctuation},\ }\href {https://doi.org/10.1103/PhysRevLett.132.036001} {\bibfield  {journal} {\bibinfo  {journal} {Phys. Rev. Lett.}\ }\textbf {\bibinfo {volume} {132}},\ \bibinfo {pages} {036001} (\bibinfo {year} {2024})}\BibitemShut {NoStop}%
\bibitem [{\citenamefont {Shavit}\ and\ \citenamefont {Alicea}(2025)}]{Alicea2025}%
  \BibitemOpen
  \bibfield  {author} {\bibinfo {author} {\bibfnamefont {G.}~\bibnamefont {Shavit}}\ and\ \bibinfo {author} {\bibfnamefont {J.}~\bibnamefont {Alicea}},\ }\bibfield  {title} {\bibinfo {title} {Quantum geometric kohn-luttinger superconductivity},\ }\href {https://doi.org/10.1103/PhysRevLett.134.176001} {\bibfield  {journal} {\bibinfo  {journal} {Phys. Rev. Lett.}\ }\textbf {\bibinfo {volume} {134}},\ \bibinfo {pages} {176001} (\bibinfo {year} {2025})}\BibitemShut {NoStop}%
\bibitem [{\citenamefont {Jahin}\ and\ \citenamefont {Lin}(2025)}]{jahin2025enhancedkohnluttingertopologicalsuperconductivity}%
  \BibitemOpen
  \bibfield  {author} {\bibinfo {author} {\bibfnamefont {A.}~\bibnamefont {Jahin}}\ and\ \bibinfo {author} {\bibfnamefont {S.-Z.}\ \bibnamefont {Lin}},\ }\href {https://arxiv.org/abs/2411.09664} {} (\bibinfo {year} {2025}),\ \Eprint {https://arxiv.org/abs/2411.09664} {arXiv:2411.09664 [cond-mat.supr-con]} \BibitemShut {NoStop}%
\bibitem [{\citenamefont {Heinsdorf}(2025)}]{heinsdorf2025altermagnetic}%
  \BibitemOpen
  \bibfield  {author} {\bibinfo {author} {\bibfnamefont {N.}~\bibnamefont {Heinsdorf}},\ }\bibfield  {title} {\bibinfo {title} {Altermagnetic instabilities from quantum geometry},\ }\href {https://doi.org/10.1103/PhysRevB.111.174407} {\bibfield  {journal} {\bibinfo  {journal} {Phys. Rev. B}\ }\textbf {\bibinfo {volume} {111}},\ \bibinfo {pages} {174407} (\bibinfo {year} {2025})}\BibitemShut {NoStop}%
\bibitem [{\citenamefont {Kudo}\ and\ \citenamefont {Yanase}(2025)}]{Kudo2025arxiv}%
  \BibitemOpen
  \bibfield  {author} {\bibinfo {author} {\bibfnamefont {K.}~\bibnamefont {Kudo}}\ and\ \bibinfo {author} {\bibfnamefont {Y.}~\bibnamefont {Yanase}},\ }\href {https://arxiv.org/abs/2505.20907} {\bibinfo {title} {Odd-parity magnetism by quantum geometry}} (\bibinfo {year} {2025}),\ \Eprint {https://arxiv.org/abs/2505.20907} {arXiv:2505.20907 [cond-mat.str-el]} \BibitemShut {NoStop}%
\bibitem [{Sug(1970)}]{Sugano1970textbook}%
  \BibitemOpen
  \bibfield  {title} {\bibinfo {title} {{Chapter II - Two Electrons in a Cubic Field}},\ }in\ \href {https://doi.org/https://doi.org/10.1016/B978-0-12-676050-7.50008-1} {\emph {\bibinfo {booktitle} {{Multiplets of Transition-Metal Ions in Crystals}}}},\ \bibinfo {series} {Pure and Applied Physics}, Vol.~\bibinfo {volume} {33},\ \bibinfo {editor} {edited by\ \bibinfo {editor} {\bibfnamefont {S.}~\bibnamefont {Sugano}}, \bibinfo {editor} {\bibfnamefont {Y.}~\bibnamefont {Tanabe}},\ and\ \bibinfo {editor} {\bibfnamefont {H.}~\bibnamefont {Kamimura}}}\ (\bibinfo  {publisher} {Elsevier},\ \bibinfo {year} {1970})\ pp.\ \bibinfo {pages} {38--65}\BibitemShut {NoStop}%
\bibitem [{\citenamefont {Watanabe}\ and\ \citenamefont {Yanase}(2018)}]{Watanabe2018prb}%
  \BibitemOpen
  \bibfield  {author} {\bibinfo {author} {\bibfnamefont {H.}~\bibnamefont {Watanabe}}\ and\ \bibinfo {author} {\bibfnamefont {Y.}~\bibnamefont {Yanase}},\ }\bibfield  {title} {\bibinfo {title} {Group-theoretical classification of multipole order: Emergent responses and candidate materials},\ }\href {https://doi.org/10.1103/PhysRevB.98.245129} {\bibfield  {journal} {\bibinfo  {journal} {Phys. Rev. B}\ }\textbf {\bibinfo {volume} {98}},\ \bibinfo {pages} {245129} (\bibinfo {year} {2018})}\BibitemShut {NoStop}%
\bibitem [{\citenamefont {Hayami}\ \emph {et~al.}(2018)\citenamefont {Hayami}, \citenamefont {Yatsushiro}, \citenamefont {Yanagi},\ and\ \citenamefont {Kusunose}}]{Hayami2018prb}%
  \BibitemOpen
  \bibfield  {author} {\bibinfo {author} {\bibfnamefont {S.}~\bibnamefont {Hayami}}, \bibinfo {author} {\bibfnamefont {M.}~\bibnamefont {Yatsushiro}}, \bibinfo {author} {\bibfnamefont {Y.}~\bibnamefont {Yanagi}},\ and\ \bibinfo {author} {\bibfnamefont {H.}~\bibnamefont {Kusunose}},\ }\bibfield  {title} {\bibinfo {title} {Classification of atomic-scale multipoles under crystallographic point groups and application to linear response tensors},\ }\href {https://doi.org/10.1103/PhysRevB.98.165110} {\bibfield  {journal} {\bibinfo  {journal} {Phys. Rev. B}\ }\textbf {\bibinfo {volume} {98}},\ \bibinfo {pages} {165110} (\bibinfo {year} {2018})}\BibitemShut {NoStop}%
\bibitem [{\citenamefont {Yatsushiro}\ \emph {et~al.}(2021)\citenamefont {Yatsushiro}, \citenamefont {Kusunose},\ and\ \citenamefont {Hayami}}]{Yatsushiro2021prb}%
  \BibitemOpen
  \bibfield  {author} {\bibinfo {author} {\bibfnamefont {M.}~\bibnamefont {Yatsushiro}}, \bibinfo {author} {\bibfnamefont {H.}~\bibnamefont {Kusunose}},\ and\ \bibinfo {author} {\bibfnamefont {S.}~\bibnamefont {Hayami}},\ }\bibfield  {title} {\bibinfo {title} {Multipole classification in 122 magnetic point groups for unified understanding of multiferroic responses and transport phenomena},\ }\href {https://doi.org/10.1103/PhysRevB.104.054412} {\bibfield  {journal} {\bibinfo  {journal} {Phys. Rev. B}\ }\textbf {\bibinfo {volume} {104}},\ \bibinfo {pages} {054412} (\bibinfo {year} {2021})}\BibitemShut {NoStop}%
\bibitem [{\citenamefont {Gao}\ \emph {et~al.}(2018)\citenamefont {Gao}, \citenamefont {Vanderbilt},\ and\ \citenamefont {Xiao}}]{Gao2018spin}%
  \BibitemOpen
  \bibfield  {author} {\bibinfo {author} {\bibfnamefont {Y.}~\bibnamefont {Gao}}, \bibinfo {author} {\bibfnamefont {D.}~\bibnamefont {Vanderbilt}},\ and\ \bibinfo {author} {\bibfnamefont {D.}~\bibnamefont {Xiao}},\ }\bibfield  {title} {\bibinfo {title} {Microscopic theory of spin toroidization in periodic crystals},\ }\href {https://doi.org/10.1103/PhysRevB.97.134423} {\bibfield  {journal} {\bibinfo  {journal} {Phys. Rev. B}\ }\textbf {\bibinfo {volume} {97}},\ \bibinfo {pages} {134423} (\bibinfo {year} {2018})}\BibitemShut {NoStop}%
\bibitem [{\citenamefont {Shitade}\ \emph {et~al.}(2018)\citenamefont {Shitade}, \citenamefont {Watanabe},\ and\ \citenamefont {Yanase}}]{Shitade2018prb}%
  \BibitemOpen
  \bibfield  {author} {\bibinfo {author} {\bibfnamefont {A.}~\bibnamefont {Shitade}}, \bibinfo {author} {\bibfnamefont {H.}~\bibnamefont {Watanabe}},\ and\ \bibinfo {author} {\bibfnamefont {Y.}~\bibnamefont {Yanase}},\ }\bibfield  {title} {\bibinfo {title} {Theory of orbital magnetic quadrupole moment and magnetoelectric susceptibility},\ }\href {https://doi.org/10.1103/PhysRevB.98.020407} {\bibfield  {journal} {\bibinfo  {journal} {Phys. Rev. B}\ }\textbf {\bibinfo {volume} {98}},\ \bibinfo {pages} {020407} (\bibinfo {year} {2018})}\BibitemShut {NoStop}%
\bibitem [{\citenamefont {Gao}\ and\ \citenamefont {Xiao}(2018)}]{Gao2018orbital}%
  \BibitemOpen
  \bibfield  {author} {\bibinfo {author} {\bibfnamefont {Y.}~\bibnamefont {Gao}}\ and\ \bibinfo {author} {\bibfnamefont {D.}~\bibnamefont {Xiao}},\ }\bibfield  {title} {\bibinfo {title} {Orbital magnetic quadrupole moment and nonlinear anomalous thermoelectric transport},\ }\href {https://doi.org/10.1103/PhysRevB.98.060402} {\bibfield  {journal} {\bibinfo  {journal} {Phys. Rev. B}\ }\textbf {\bibinfo {volume} {98}},\ \bibinfo {pages} {060402} (\bibinfo {year} {2018})}\BibitemShut {NoStop}%
\bibitem [{\citenamefont {Shitade}\ \emph {et~al.}(2019)\citenamefont {Shitade}, \citenamefont {Daido},\ and\ \citenamefont {Yanase}}]{Shitade2019prb}%
  \BibitemOpen
  \bibfield  {author} {\bibinfo {author} {\bibfnamefont {A.}~\bibnamefont {Shitade}}, \bibinfo {author} {\bibfnamefont {A.}~\bibnamefont {Daido}},\ and\ \bibinfo {author} {\bibfnamefont {Y.}~\bibnamefont {Yanase}},\ }\bibfield  {title} {\bibinfo {title} {Theory of spin magnetic quadrupole moment and temperature-gradient-induced magnetization},\ }\href {https://doi.org/10.1103/PhysRevB.99.024404} {\bibfield  {journal} {\bibinfo  {journal} {Phys. Rev. B}\ }\textbf {\bibinfo {volume} {99}},\ \bibinfo {pages} {024404} (\bibinfo {year} {2019})}\BibitemShut {NoStop}%
\bibitem [{\citenamefont {Daido}\ \emph {et~al.}(2020)\citenamefont {Daido}, \citenamefont {Shitade},\ and\ \citenamefont {Yanase}}]{Daido2020prb}%
  \BibitemOpen
  \bibfield  {author} {\bibinfo {author} {\bibfnamefont {A.}~\bibnamefont {Daido}}, \bibinfo {author} {\bibfnamefont {A.}~\bibnamefont {Shitade}},\ and\ \bibinfo {author} {\bibfnamefont {Y.}~\bibnamefont {Yanase}},\ }\bibfield  {title} {\bibinfo {title} {Thermodynamic approach to electric quadrupole moments},\ }\href {https://doi.org/10.1103/PhysRevB.102.235149} {\bibfield  {journal} {\bibinfo  {journal} {Phys. Rev. B}\ }\textbf {\bibinfo {volume} {102}},\ \bibinfo {pages} {235149} (\bibinfo {year} {2020})}\BibitemShut {NoStop}%
\bibitem [{\citenamefont {Kitamura}\ \emph {et~al.}(2021)\citenamefont {Kitamura}, \citenamefont {Ishizuka}, \citenamefont {Daido},\ and\ \citenamefont {Yanase}}]{Kitamura2021prb}%
  \BibitemOpen
  \bibfield  {author} {\bibinfo {author} {\bibfnamefont {T.}~\bibnamefont {Kitamura}}, \bibinfo {author} {\bibfnamefont {J.}~\bibnamefont {Ishizuka}}, \bibinfo {author} {\bibfnamefont {A.}~\bibnamefont {Daido}},\ and\ \bibinfo {author} {\bibfnamefont {Y.}~\bibnamefont {Yanase}},\ }\bibfield  {title} {\bibinfo {title} {Thermodynamic electric quadrupole moments of nematic phases from first-principles calculations},\ }\href {https://doi.org/10.1103/PhysRevB.103.245114} {\bibfield  {journal} {\bibinfo  {journal} {Phys. Rev. B}\ }\textbf {\bibinfo {volume} {103}},\ \bibinfo {pages} {245114} (\bibinfo {year} {2021})}\BibitemShut {NoStop}%
\bibitem [{\citenamefont {Nogaki}\ and\ \citenamefont {Yanase}(2024)}]{Nogaki2024prb}%
  \BibitemOpen
  \bibfield  {author} {\bibinfo {author} {\bibfnamefont {K.}~\bibnamefont {Nogaki}}\ and\ \bibinfo {author} {\bibfnamefont {Y.}~\bibnamefont {Yanase}},\ }\bibfield  {title} {\bibinfo {title} {Field-induced superconductivity mediated by odd-parity multipole fluctuation},\ }\href {https://doi.org/10.1103/PhysRevB.110.184501} {\bibfield  {journal} {\bibinfo  {journal} {Phys. Rev. B}\ }\textbf {\bibinfo {volume} {110}},\ \bibinfo {pages} {184501} (\bibinfo {year} {2024})}\BibitemShut {NoStop}%
\bibitem [{\citenamefont {Cavanagh}\ \emph {et~al.}(2022)\citenamefont {Cavanagh}, \citenamefont {Shishidou}, \citenamefont {Weinert}, \citenamefont {Brydon},\ and\ \citenamefont {Agterberg}}]{Cavanagh2022prb}%
  \BibitemOpen
  \bibfield  {author} {\bibinfo {author} {\bibfnamefont {D.~C.}\ \bibnamefont {Cavanagh}}, \bibinfo {author} {\bibfnamefont {T.}~\bibnamefont {Shishidou}}, \bibinfo {author} {\bibfnamefont {M.}~\bibnamefont {Weinert}}, \bibinfo {author} {\bibfnamefont {P.~M.~R.}\ \bibnamefont {Brydon}},\ and\ \bibinfo {author} {\bibfnamefont {D.~F.}\ \bibnamefont {Agterberg}},\ }\bibfield  {title} {\bibinfo {title} {Nonsymmorphic symmetry and field-driven odd-parity pairing in $\mathrm{Ce}{\mathrm{rh}}_{2}{\mathrm{as}}_{2}$},\ }\href {https://doi.org/10.1103/PhysRevB.105.L020505} {\bibfield  {journal} {\bibinfo  {journal} {Phys. Rev. B}\ }\textbf {\bibinfo {volume} {105}},\ \bibinfo {pages} {L020505} (\bibinfo {year} {2022})}\BibitemShut {NoStop}%
\bibitem [{\citenamefont {Suh}\ \emph {et~al.}(2023)\citenamefont {Suh}, \citenamefont {Yu}, \citenamefont {Shishidou}, \citenamefont {Weinert}, \citenamefont {Brydon},\ and\ \citenamefont {Agterberg}}]{Suh2023prr}%
  \BibitemOpen
  \bibfield  {author} {\bibinfo {author} {\bibfnamefont {H.~G.}\ \bibnamefont {Suh}}, \bibinfo {author} {\bibfnamefont {Y.}~\bibnamefont {Yu}}, \bibinfo {author} {\bibfnamefont {T.}~\bibnamefont {Shishidou}}, \bibinfo {author} {\bibfnamefont {M.}~\bibnamefont {Weinert}}, \bibinfo {author} {\bibfnamefont {P.~M.~R.}\ \bibnamefont {Brydon}},\ and\ \bibinfo {author} {\bibfnamefont {D.~F.}\ \bibnamefont {Agterberg}},\ }\bibfield  {title} {\bibinfo {title} {Superconductivity of anomalous pseudospin in nonsymmorphic materials},\ }\href {https://doi.org/10.1103/PhysRevResearch.5.033204} {\bibfield  {journal} {\bibinfo  {journal} {Phys. Rev. Res.}\ }\textbf {\bibinfo {volume} {5}},\ \bibinfo {pages} {033204} (\bibinfo {year} {2023})}\BibitemShut {NoStop}%
\bibitem [{\citenamefont {Sukhachov}\ \emph {et~al.}(2025)\citenamefont {Sukhachov}, \citenamefont {Aase}, \citenamefont {M\ae{}land},\ and\ \citenamefont {Sudb\o{}}}]{Sukhachov2025prb}%
  \BibitemOpen
  \bibfield  {author} {\bibinfo {author} {\bibfnamefont {P.}~\bibnamefont {Sukhachov}}, \bibinfo {author} {\bibfnamefont {N.~H.}\ \bibnamefont {Aase}}, \bibinfo {author} {\bibfnamefont {K.}~\bibnamefont {M\ae{}land}},\ and\ \bibinfo {author} {\bibfnamefont {A.}~\bibnamefont {Sudb\o{}}},\ }\bibfield  {title} {\bibinfo {title} {Effect of the hubbard interaction on the quantum metric},\ }\href {https://doi.org/10.1103/PhysRevB.111.085143} {\bibfield  {journal} {\bibinfo  {journal} {Phys. Rev. B}\ }\textbf {\bibinfo {volume} {111}},\ \bibinfo {pages} {085143} (\bibinfo {year} {2025})}\BibitemShut {NoStop}%
\bibitem [{\citenamefont {Momma}\ and\ \citenamefont {Izumi}(2011)}]{Momma2011jac}%
  \BibitemOpen
  \bibfield  {author} {\bibinfo {author} {\bibfnamefont {K.}~\bibnamefont {Momma}}\ and\ \bibinfo {author} {\bibfnamefont {F.}~\bibnamefont {Izumi}},\ }\bibfield  {title} {\bibinfo {title} {{{\it VESTA3} for three-dimensional visualization of crystal, volumetric and morphology data}},\ }\href {https://doi.org/10.1107/S0021889811038970} {\bibfield  {journal} {\bibinfo  {journal} {Journal of Applied Crystallography}\ }\textbf {\bibinfo {volume} {44}},\ \bibinfo {pages} {1272} (\bibinfo {year} {2011})}\BibitemShut {NoStop}%
\bibitem [{\citenamefont {Kokalj}(1999)}]{Kokaj1999jmgm}%
  \BibitemOpen
  \bibfield  {author} {\bibinfo {author} {\bibfnamefont {A.}~\bibnamefont {Kokalj}},\ }\bibfield  {title} {\bibinfo {title} {Xcrysden—a new program for displaying crystalline structures and electron densities},\ }\href {https://doi.org/https://doi.org/10.1016/S1093-3263(99)00028-5} {\bibfield  {journal} {\bibinfo  {journal} {Journal of Molecular Graphics and Modelling}\ }\textbf {\bibinfo {volume} {17}},\ \bibinfo {pages} {176} (\bibinfo {year} {1999})}\BibitemShut {NoStop}%
\bibitem [{\citenamefont {Kawamura}(2019)}]{Kawamura2019cpc}%
  \BibitemOpen
  \bibfield  {author} {\bibinfo {author} {\bibfnamefont {M.}~\bibnamefont {Kawamura}},\ }\bibfield  {title} {\bibinfo {title} {Fermisurfer: Fermi-surface viewer providing multiple representation schemes},\ }\href {https://doi.org/https://doi.org/10.1016/j.cpc.2019.01.017} {\bibfield  {journal} {\bibinfo  {journal} {Computer Physics Communications}\ }\textbf {\bibinfo {volume} {239}},\ \bibinfo {pages} {197} (\bibinfo {year} {2019})}\BibitemShut {NoStop}%
\bibitem [{\citenamefont {Luttinger}\ and\ \citenamefont {Ward}(1960)}]{Luttinger1960pr}%
  \BibitemOpen
  \bibfield  {author} {\bibinfo {author} {\bibfnamefont {J.~M.}\ \bibnamefont {Luttinger}}\ and\ \bibinfo {author} {\bibfnamefont {J.~C.}\ \bibnamefont {Ward}},\ }\bibfield  {title} {\bibinfo {title} {{Ground-State Energy of a Many-Fermion System. II}},\ }\href {https://doi.org/10.1103/PhysRev.118.1417} {\bibfield  {journal} {\bibinfo  {journal} {Phys. Rev.}\ }\textbf {\bibinfo {volume} {118}},\ \bibinfo {pages} {1417} (\bibinfo {year} {1960})}\BibitemShut {NoStop}%
\bibitem [{\citenamefont {Luttinger}(1960)}]{Luttinger1960pr2}%
  \BibitemOpen
  \bibfield  {author} {\bibinfo {author} {\bibfnamefont {J.~M.}\ \bibnamefont {Luttinger}},\ }\bibfield  {title} {\bibinfo {title} {{Fermi Surface and Some Simple Equilibrium Properties of a System of Interacting Fermions}},\ }\href {https://doi.org/10.1103/PhysRev.119.1153} {\bibfield  {journal} {\bibinfo  {journal} {Phys. Rev.}\ }\textbf {\bibinfo {volume} {119}},\ \bibinfo {pages} {1153} (\bibinfo {year} {1960})}\BibitemShut {NoStop}%
\bibitem [{\citenamefont {Baym}\ and\ \citenamefont {Kadanoff}(1961)}]{Baym1961pr}%
  \BibitemOpen
  \bibfield  {author} {\bibinfo {author} {\bibfnamefont {G.}~\bibnamefont {Baym}}\ and\ \bibinfo {author} {\bibfnamefont {L.~P.}\ \bibnamefont {Kadanoff}},\ }\bibfield  {title} {\bibinfo {title} {{Conservation Laws and Correlation Functions}},\ }\href {https://doi.org/10.1103/PhysRev.124.287} {\bibfield  {journal} {\bibinfo  {journal} {Phys. Rev.}\ }\textbf {\bibinfo {volume} {124}},\ \bibinfo {pages} {287} (\bibinfo {year} {1961})}\BibitemShut {NoStop}%
\bibitem [{\citenamefont {Baym}(1962)}]{Baym1962pr}%
  \BibitemOpen
  \bibfield  {author} {\bibinfo {author} {\bibfnamefont {G.}~\bibnamefont {Baym}},\ }\bibfield  {title} {\bibinfo {title} {{Self-Consistent Approximations in Many-Body Systems}},\ }\href {https://doi.org/10.1103/PhysRev.127.1391} {\bibfield  {journal} {\bibinfo  {journal} {Phys. Rev.}\ }\textbf {\bibinfo {volume} {127}},\ \bibinfo {pages} {1391} (\bibinfo {year} {1962})}\BibitemShut {NoStop}%
\end{thebibliography}%

\end{document}